\documentclass[a4paper,11pt]{article}

\usepackage[english]{babel}
\usepackage[utf8x]{inputenc}
\usepackage{siunitx}
\usepackage{booktabs}
\usepackage{comment}

\usepackage{amsmath,amssymb}
\usepackage{graphicx}
\usepackage[colorlinks=true, allcolors=black]{hyperref}
\usepackage{geometry}
\usepackage{natbib}
\usepackage{graphics}
\usepackage{setspace}

\begin{document}

\title{The Economics of $p(doom)$: Scenarios of Existential Risk and Economic Growth in the Age of Transformative AI\thanks{We would like to thank two anonymous Reviewers, the Editor Angus Chu, the Associate Editor, as well as Ben Jones, Chad Jones, Torben Klarl, Anton Korinek, Anna Kortis, and Philip Trammell for helpful comments and suggestions. We acknowledge helpful discussions at the 1st CEPR Workshop of the Research and Policy Network on AI (Paris, December 2024) and the Joint NBP-SNB Seminar (Warsaw, October 2025). We are grateful for the research funding from the National Science Centre Poland through the grant ``Will Artificial General Intelligence Bring Extinction or Cornucopia? Modeling the Economy at Technological Singularity'' (OPUS 26 No. 2023/51/B/HS4/00096). No generative AI tools were used in the preparation of the manuscript. All errors are our responsibility.}}


\author{Jakub Growiec\thanks{SGH Warsaw School of Economics, Poland and CEPR Research and Policy Network on AI.} \and Klaus Prettner\thanks{Vienna University of Economics and Business (WU), Austria.}}

\maketitle



\begin{abstract}

Recent advances in artificial intelligence (AI) have led to a wide range of predictions about its long-term impact on humanity. A central focus is the potential emergence of transformative AI (TAI), eventually capable of outperforming humans in all economically valuable tasks and fully automating labor. Discussed scenarios range from unprecedented economic growth and abundance (“post-scarcity'' or ``cornucopia'') to human extinction after a misaligned TAI takes over (“AI doom''). However, the probabilities and implications of these scenarios remain highly uncertain. We contribute by organizing the various scenarios and evaluating their associated existential risks and economic outcomes in terms of aggregate welfare. Our results imply that even low-probability catastrophic outcomes justify substantial investments in AI safety and alignment research. This result highlights that current global efforts in AI safety and alignment research are insufficient relative to the scale and urgency of the risks posed by TAI.
\vspace{0.2mm} \\
\textbf{JEL codes:} I30, O11, O33, Q01. \\
\textbf{Keywords:} Transformative Artificial Intelligence (TAI), Economic Growth, Technological Singularity, Growth Explosion, AI Takeover, AI Alignment, AI Doom. 
\end{abstract}

\newpage

\section{Introduction}
Recent breakthroughs in the area of artificial intelligence (AI) have led to a wide range of predictions for the future of humankind. Of particular interest is the expected future arrival of transformative artificial intelligence (TAI), allowing unaided machines to perform every economically valuable task better than humans and thereby, if implemented, fully automate human labor. Following such an event, ``techno-optimists'' anticipate an explosion of productivity and economic growth that will empower humanity to achieve unimaginable wealth (sometimes referred to as ``cornucopia'' or ``post-scarcity''). By contrast, ``AI doomers'' expect that superintelligent TAI will take over the world's key decision making processes, which will likely lead to permanent disempowerment of humanity or even human extinction.\footnote{In this context, individual estimates of $p(doom)$, referring typically to the probability of human extinction within the 21st century, are often invoked. Specifically, $p(doom|TAI)$ refers to human extinction following the arrival of superintelligent, transformative AI, broadly agreed to be the prime contributor to $p(doom)$.} Some predictions, often from economists, occupy a middle ground, claiming that 
both the fears of human extinction and hopes for growth explosion are hugely exaggerated 
\citep[see, for example,][for the very different views]{Bostrom2014, Pratt2015, Tegmark2017, Ord2020, nordhausAreWeApproaching2021, Yudkowsky2022, Allyn-Feuer2023, Amodei, Amodei2026, BengioEtal2024, Skare2024, Jones2023, Jones2025, Acemoglu2024Thesimple, IABIED, Uddin2026}.


Given this wide range of possible outcomes, ranging from cornucopia to human extinction, our aim is to contribute by (i) systematically characterizing the possible future pathways of humanity in the age of TAI; (ii) assessing their value from a human-centric social welfare perspective;
and (iii) arguing for the importance of reducing the likelihood of catastrophic outcomes through AI safety and AI alignment research.
In our normative study we ask two key research questions: how much existential risk would a benevolent social planner tolerate in exchange for accelerated economic growth after an AI takeover, and how much would the planner be willing to pay to avoid that risk?


We formalize a number of scenarios of existential risk and economic growth, with a special focus on scenarios involving an AI takeover. 
Our baseline is a (vastly) positive scenario of ``cornucopia'' in which a benevolent TAI, whose goals are perfectly aligned with human flourishing, maximizes long-run human utility in the context of a fully automated economy growing at a rate proportional to the growth rate of programmable hardware such as compute and robots \citep{Prettner2017,GrowiecEtal2024}. Against this scenario, we describe a number of failure modes arising due to TAI misalignment. These scenarios involve existential risk, e.g., in the form of human extinction---either immediately after the AI takeover or at a later date.\footnote{Existential risks are ``risks that threaten the destruction of humanity's long-term potential'' \citep{Ord2020}. They include the risk of human extinction as well as scenarios in which humans survive but are irreversibly locked in a drastically inferior state of affairs and can no longer develop a civilization.} Sufficiently powerful misaligned TAI may imply immediate human extinction \citep{Jones2023}. But extinction may also occur later, either due to a random event realizing the extinction risk that had been quietly mounting in the background or due to the actions of the misaligned TAI after it achieves new information or a new capability. 




The key contribution of this paper to the literature is a quantitative assessment of the different future scenarios, with and without TAI, from a social welfare perspective. We find that it is worth investing in reducing the risk of AI doom (human extinction or other catastrophic outcomes) even if the risk of them occurring were low.\footnote{Because of many suggestions that the risk may actually be much higher \citep[see e.g.,][]{Bostrom2014,Dung2024,Yampolskiy2024,IABIED}, our results are on the conservative side.} Specifically, we find that a benevolent social planner, acting optimally on behalf of humankind, would be willing to pay high amounts to prevent extinction risk. In the cases where the extinction risk cannot be reduced to zero, the social planner may even prefer to refrain from the development of TAI. This baseline result, obtained under the assumption of a realistic level of risk aversion, can be overturned only if one allows for unbounded flow utility, in which case the utility gain achieved in cornucopia can, in expectation, outweigh the loss of future utility after human extinction \citep{Jones2023}. But even in such a counterfactual case, the social planner exhibits a remarkable willingness to pay for interventions that reduce existential risk.  
Our results are robust to a wide variety of parameter choices, underscoring the generality of our main conclusion that currently there is an alarming underinvestment in AI safety and AI alignment research, leading to humanity's excessive exposure to existential risk. 

This alarming conclusion could be further strengthened by observing that our focus on scenarios of AI takeover\footnote{AI takeover may occur due to humans voluntarily giving up (parts of) their decision power to improve efficiency, enable economic expansion, and withstand competitive pressures; or it could also be because TAI may pursue its own goals, potentially conflicting with ours. It is important to note that AI takeover does not require any ``consciousness'' or ``hostility'' of TAI but only a conflict between its objectives and the ones of humans (e.g., over allocating energy, computing power, etc.).} 
ignores additional channels through which TAI can bring  adverse outcomes, such as through accidents or deliberate misuses by malevolent human actors \citep[see, for example,][]{Burton2023,BengioEtal2024}. Thus, the total risk from TAI should be considered even greater than what our analysis takes into account.

The article is structured as follows. In Section \ref{sec:motivation}, we motivate our research based on a multidisciplinary review of the literature. In Section \ref{sec:taxonomy}, we provide a systematic discussion of the different possible outcomes of AI development for humanity and their probabilities. In Section \ref{sec:modelling}, we set up a formal model of the different scenarios. In Section \ref{sec:modelofpdoom}, we compare the outcomes from a welfare perspective and show that investing in AI safety and AI alignment research is imperative. Section \ref{sec:concl} concludes.

\section{Motivation and Literature Review}
\label{sec:motivation}

In this section, we provide a multidisciplinary review of the literature and clarify the main concepts and definitions that matter in the context of TAI outcomes and safety.

\subsection{AI as General Purpose Technology}

AI models, which may take the form of, e.g., predictive algorithms, chatbots, copilots, or software agents, are General Purpose Technologies: like the steam engine, electricity or computers, they are applicable across all economic sectors and are ``characterized by pervasiveness, inherent potential for technical improvements, and `innovational complementarities', giving rise to increasing returns-to-scale'' \citep[][p. 83]{BresnahanTrajtenberg1995}. Hence, they offer sizeable potential for long-run growth acceleration, even if their adoption may require complementary investments needed to rearrange corporate workflows and organizational structures \citep{AgrawalEtal2022}, and may involve a prolonged transition period during which aggregate productivity stagnates or even declines \citep[along a ``J-curve'',][]{BrynjolfssonEtal2017,GPTs}.

The adoption of AI in the economy has been frequently modeled in the growth literature as task automation \citep[see e.g.,][]{AcemogluRestrepo2018,Acemoglu2024Thesimple}, similarly to the use of industrial robots \citep{DeCanio2016,GraetzMichaels2018}. Viewed through such a lens, AI's potential for growth acceleration may appear limited. However, the logic of gradual task replacement breaks down when AI is also used in research, which creates additional feedback loops \citep{AJJ-AI,BesirogluEtal2022,DavidsonEtal2026}, or when AI exceeds human capability across the entire distribution of task complexity, allowing for full automation of human cognitive work \citep{KorinekSuh2024,GrowiecEtal2024,CataliniEtal2026}. This is the perspective taken by the economics of transformative AI.

\subsection{Transformative AI and Technological Singularity}

By \emph{transformative AI} (TAI) we understand a suite of AI algorithms allowing unaided machines to perform every economically valuable task better than humans and thereby, if implemented, fully automate human labor. In the words of \cite{Karnofsky2016}, TAI would be an ``AI that precipitates a transition comparable to (or more significant than) the agricultural or industrial revolution.'' The concept of TAI largely coincides with \emph{artificial general intelligence} (AGI), defined usually as ``a type of AI that matches or surpasses human cognitive capabilities across a wide range of cognitive tasks.'' 
However, because of the additional requirement of wide economic applicability, the bar for TAI is somewhat higher than for AGI. Accordingly, while some first prominent voices are arguing that AGI is already here \citep{Norvig2023,Nature_AGI}, transformative economic impacts of this technology are yet to come.  

All in all, while the exact definition of TAI may be somewhat fuzzy,
there is a growing consensus that TAI is approaching fast. For example,  the following years have been stated as the expected arrival dates of TAI by different sources: 2027 \citep{Aschenbrenner2024,AI2027}, 2032 (metaculus.com median forecast as of April 2026), 2033 \citep[the ``direct approach'' model by][]{EpochAIDirect}, 2040 \citep[the ``bio anchors'' model by][]{Cotra2022}, and 2047 \citep[according to a survey of AI researchers,][]{GraceEtal2024}.\footnote{Of course, timelines to TAI are uncertain. For example, \cite{Allyn-Feuer2023} provide a counterargument to short timelines.} Once developed, TAI will be highly agentic, able to control a variety of physical actuators, and possibly able to improve itself through a cascade of recursive self-improvements to become decidedly superhuman (or equivalently, to develop its superhuman successor). This ``intelligence explosion'' (or ``take-off'') phase may 
culminate in \emph{technological singularity} \citep[as described and discussed by][]{Kurzweil2005,Roodman2020,Davidson2023}---a state in which human input will no longer be needed to sustain the global economy and civilization.

The arrival of TAI would be the threshold moment for technological singularity because as all essential production and R\&D tasks become fully automatable, people and machines would switch from being complements to substitutes \citep{Growiec2020}. When human cognitive work is no longer essential for production, people will only find employment as long as they are price-competitive against the machines---a position in which people are destined to lose eventually. On the other hand, once the human input  ceases to be the bottleneck of economic growth, the key growth engine would no longer be labor-augmenting technological change, as it has been throughout the 20th century, but rather the accumulation of compute and robots \citep{Prettner2017, Growiec2022Book, GrowiecEtal2024}. Output growth rates could then rise towards the pace of compute accumulation, which follows the rate of Moore's Law (historically 20-30\% per annum)---i.e., an order of magnitude faster than historical growth in global GDP. 

 To see this from a theoretical perspective, consider the following example. In any model of capital--labor automation with constant elasticity of substitution (CES) production, under gross complementarity between capital and labor (elasticity of substitution between 0 and 1), long-run growth is not possible unless there is also labor-augmenting technological progress. The same result follows for the Cobb-Douglas case (with a unitary elasticity of substitution). By contrast, under gross substitutability, long-run growth without technological progress (i.e., through capital accumulation alone) is perfectly possible \citep[][]{Steigum2011, Prettner2017}. Depending on model parameters, the threshold of the elasticity of substitution may lie somewhere in the gross substitutability space between 1 and $+\infty$ \citep[][]{Lankisch2019} but perfect substitutability is not necessary for endogenous growth via capital accumulation only. Hence, in the context of TAI the growth takeoff obtains from the fact that unlike human brains, physical capital---and, specifically, digital compute able to run the AI software and thereby perform all the essential cognitive tasks---is accumulable per capita.

Although a strict mathematical singularity, according to which output reaches infinity in finite time, is not possible physically, a strong acceleration of economic growth could occur with TAI \citep[][]{TrammellKorinek2020,Davidson2021}. There are both downside and upside risks to this prediction. On the downside, growth in AI capabilities could be slowed down by physical constraints, such as energy availability, quality of the electric grid, cooling requirements, limits to chip miniaturization, etc. Yet, some of these constraints can be resolved or circumvented with sufficient investment---therefore, in our analysis, we assume that like in the Solow or AK models, energy and other hardware complementary to compute is sufficiently available so that it will not bottleneck economic growth. On the upside, accelerated scientific progress thanks to TAI could also possibly lead to new breakthroughs in energy production or computing efficiency, for example, through fusion power or quantum computing, thereby potentially accelerating Moore's Law beyond historical rates. Because all these possibilities are highly speculative, we stick to our baseline expectation that even after the arrival of TAI, Moore's Law will advance at rates similar to historical ones.\footnote{Note that we are expecting economic growth to catch up with the growth of general purpose compute and robots rather than specifically training compute and algorithmic efficiency of frontier AI models, which has been recently growing much faster than Moore's Law \citep{epoch2023aitrends}.}

\subsection{AI Takeover}


There are two distinct reasons to expect some form of AI takeover of key decision making processes once TAI is developed. First, there is the economic rationale: in any decentralized setting, people have an incentive to empower AI to cut costs, improve productivity, gain market share, and increase individual utility. Also, while each individual, firm, or other formal entity may be willing to retain at least some discretion in their decision making, they may be forced to abandon it in favor of more comprehensive, efficiency-enhancing AI automation due to cut-throat competitive pressures from those who do not impose such constraints.\footnote{The pressure to adopt AI-powered decision making is perhaps strongest in military applications. Already today AI is increasingly used in warfare, e.g., to identify targets and to steer autonomous drones. 
} 
These effects can aggregate up to the level of the entire economy: keeping inefficient, orders of magnitude slower human decision making in the loop would bottleneck economic growth severely \citep{Growiec2023GN,Growiec2022XS2} and preclude technological singularity. The lingering presence of unutilized technological opportunity, which could potentially be unleashed by motivated actors, would make such a scenario unsustainable over a longer time frame \citep{Growiec2022Book,Dung2024,Yampolskiy2024}.

Second, there is the technological rationale. 
Namely, TAI is expected to attempt a takeover because of the way AI is currently developed \citep{Cotra2022Takeover}. For example, AI models trained with human feedback on diverse tasks are by default expected to become competent, creative, forward-looking agents ``playing the training game'' to maximize reward by any means. Also, due to instrumental convergence \citep{Bostrom2014, Turner2023, Tarsney2025}, which we discuss in more detail below, TAI will strive to self-preserve and acquire resources to achieve its goals. It may claim the decision-making authority following a the conflict between humans and TAI over Earth's resources such as energy and raw materials. 



An unappealing feature of AI takeover is a strong force toward the centralization of decision-making. This expectation is based on the scaling properties of digital software. Namely, digital code can be costlessly, perfectly copied; clones and subroutines of the master AI algorithm can then be executed at the scale of all available hardware. This is different from the human population where people's cognitive powers are bound to their brains which cannot be copied across individuals or devices. Hence, even in the multipolar scenario in which multiple TAIs are built simultaneously, differences in intelligence across AI architectures would likely compound through compute scaling and recursive self-improvement, and eventually one TAI may dominate the others.\footnote{
Note that a multipolar scenario in which the balance of power between multiple competing TAIs is sustained for a prolonged time period fares badly for human empowerment and access to resources, too. See also \cite{Yudkowsky2013, KorinekStiglitz2019}.} 

Last but not least, due to the systematic, sizable discrepancy in growth rates between (technology-augmented) human cognitive powers and AI capabilities, we expect the AI takeover to be irreversible.


\subsection{TAI Alignment}

TAI alignment---i.e., the compatibility of TAI goals with long-run flourishing of humankind---will be necessary for technological singularity to benefit humans.\footnote{Throughout the paper, by TAI alignment we mean the alignment  of goals and values of agentic AIs with long-run flourishing of humankind. This is different from ``alignment'' in the context of, for example, today's large language models, consisting in gradually reducing the incidence  of unwanted outputs through reinforcement learning from human feedback (RLHF). There is a consensus that RLHF is not scalable and therefore it will not work for TAI \citep{Aschenbrenner2024,Hubinger2025}. } Upon AI takeover, humankind loses the ability to interfere with TAI's actions, which makes it necessary to ensure, prior to developing the TAI, that it will not perform actions that could harm us. While so critical, achieving TAI alignment is also hard, for at least six reasons. First, the \emph{orthogonality principle} \citep{Bostrom2014}: any level of intelligence, or optimization power, can be coupled with any final goal. It should not be expected that intelligence itself will ensure AI alignment by default. 

Second, the \emph{instrumental convergence thesis} \citep{Bostrom2014, Turner2023, Tarsney2025}: any open-ended 
final goal implies that the TAI will also follow the following four instrumental goals to achieve its final goal: (i) self-preservation and goal integrity, (ii) resource acquisition, (iii) efficiency, and (iv) curiosity / technological perfection. Pursuit of these emergent instrumental goals implies that the TAI will seek power and control over resources.

The third reason is Goodhart's Law: ``any observed statistical regularity will tend to collapse once pressure is placed upon it for control purposes'' \citep{ManheimGarrabrant2019}. Humanity does not know its goals and values exactly, particularly in reference to futures involving radically different, superior technology. As our civilization develops, we gradually learn about the side effects of our actions and update our beliefs on the desired future actions and states. This makes us chase imperfect proxies of our species' long-run flourishing, such as GDP per capita or reported life satisfaction. Unfortunately, this is a major challenge to \emph{outer alignment}, i.e., the problem of representing the goals that the TAI should pursue. Like in the myth of King Midas, we ought to be very careful what we are wishing for, because we may as well get it \citep{IABIED}. Overall, with Goodhart's Law comes the ``Anna Karenina principle'': there is only one way to get outer alignment right, and multiple ways to get it wrong---i.e., there are many failure modes.

Fourth, \emph{inner alignment} is also going to be a challenge: due to unforeseen emergent forces, final TAI goals may deviate from the goals specified in the training process \citep{Bostrom2014}. Its goals may also be unstable or corruptible as the TAI develops further or learns new important facts about its environment. For example, it may find it worthwhile to reshape the world economy into a ``self-contained `production web' of companies operating with no human involvement, (...) [able to] decouple economic activities from serving human values'' \citep[][p. 5]{CritchRussell2023}.

Fifth, there is no room for trial and error. Logically, the TAI alignment problem must be resolved successfully already on the first critical try \citep{IABIED}---after it successfully assumes control over critical decision-making processes, owing to its instrumental goals, it will not allow itself to be switched off or be reprogrammed.

Sixth, short timelines to TAI and the race dynamics in the modern-day AI industry \citep{epoch2023aitrends} make the AI alignment research program very time constrained. There is also no clear consensus in the industry regarding the importance and appropriate approach to value alignment, and regulation only comes late and is still predominantly focused on privacy and intellectual property rights, not on lowering existential risks.\footnote{Race dynamics are similar to a typical Nash equilibrium situation in which no credible coordination is possible to get to the outcome with maximum utility. Even if it could be solved within one country, for example the USA, if China and other actors do not cooperate, it would also fail.} These circumstances limit the chances of success. 

Thus far, the industry's main strategy for TAI alignment is to ``muddle through'' the problem \citep{Aschenbrenner2024} via scalable oversight, that is, by creating a sequence of consecutive generations of automated AI researchers able to perform reinforcement learning techniques on next-generation AI algorithms of incrementally increasing capability. This was, in a nutshell, the vision put forward by the now dismantled OpenAI Superalignment team led by Jan Leike and Ilya Sutskever\footnote{\url{https://openai.com/index/introducing-superalignment/} [access: 23.12.2024]. See also \cite{Hubinger2025} for Anthropic's perspective.}. But there are also voices that with the current machine learning approach, TAI alignment is not feasible, and focus should instead be shifted to human-compatible \citep{Russell2019}, or provably safe AI systems \citep{TegmarkOmohundro2023}. 



\subsection{Corrigibility}

Assuming that when the TAI takes over, it has well-aligned goals and demonstrably cares for the long-run flourishing of humankind, can we be sure that it will remain ``friendly'' forever? Unfortunately not, for two reasons. First, over time, we may learn new facts about the world, suggesting that our initial view of alignment was flawed and should be adjusted. This is particularly likely if the TAI's actions lead to deep changes in human living conditions and the Earth's environment in general.
After all, our understanding of human needs and values has repeatedly changed in the past driven by technological and scientific progress, human and social development, and economic growth. It is almost certain that it will also change in the future, particularly such a transformative one.

Second, the TAI may undergo a sequence of transformations or be replaced by a new, self-made, even more powerful successor. Differently to the human brain, TAI software (its learning algorithm) will likely be improving over time. However, during the updating process its goals may be re-interpreted and some of their implications may change. Specifically, they may deviate from long-run human flourishing.

This is the problem of \emph{corrigibility} \citep{SoaresEtal2015}. To avoid AI doom, TAI goals must not only be well-aligned, but also corrigible: adapting to new information and stable under TAI self-improvements. The TAI must not just have an adequate, but also a flexible representation of human utility (or welfare), allowing for recursive modifications of the utility specification as new information is learned, while preventing inadequate updates that could turn out deadly.

Unfortunately, in light of the instrumental convergence thesis, the default outcome is that the values are permanently locked in. This is because any hypothetical change in values would run counter to the present values that are being optimized for, so the algorithm will naturally resist such change.  \cite{GreenblattEtal2024} have observed this behavior in a present-day large language model, Anthropic's Claude.

Addressing the additional issue of corrigibility and---more broadly---allowing for the possibility that the existential risk from TAI may materialize with a delay, distinguishes our approach from that of \cite{Jones2023,Jones2025} who considers the development of TAI as a one-off risk.


\section{A Taxonomy of Outcomes for Humanity}
\label{sec:taxonomy}

\subsection{Overview of Scenarios}

\begin{figure}[!h]
	\begin{center}
		\includegraphics[width=0.99\textwidth]{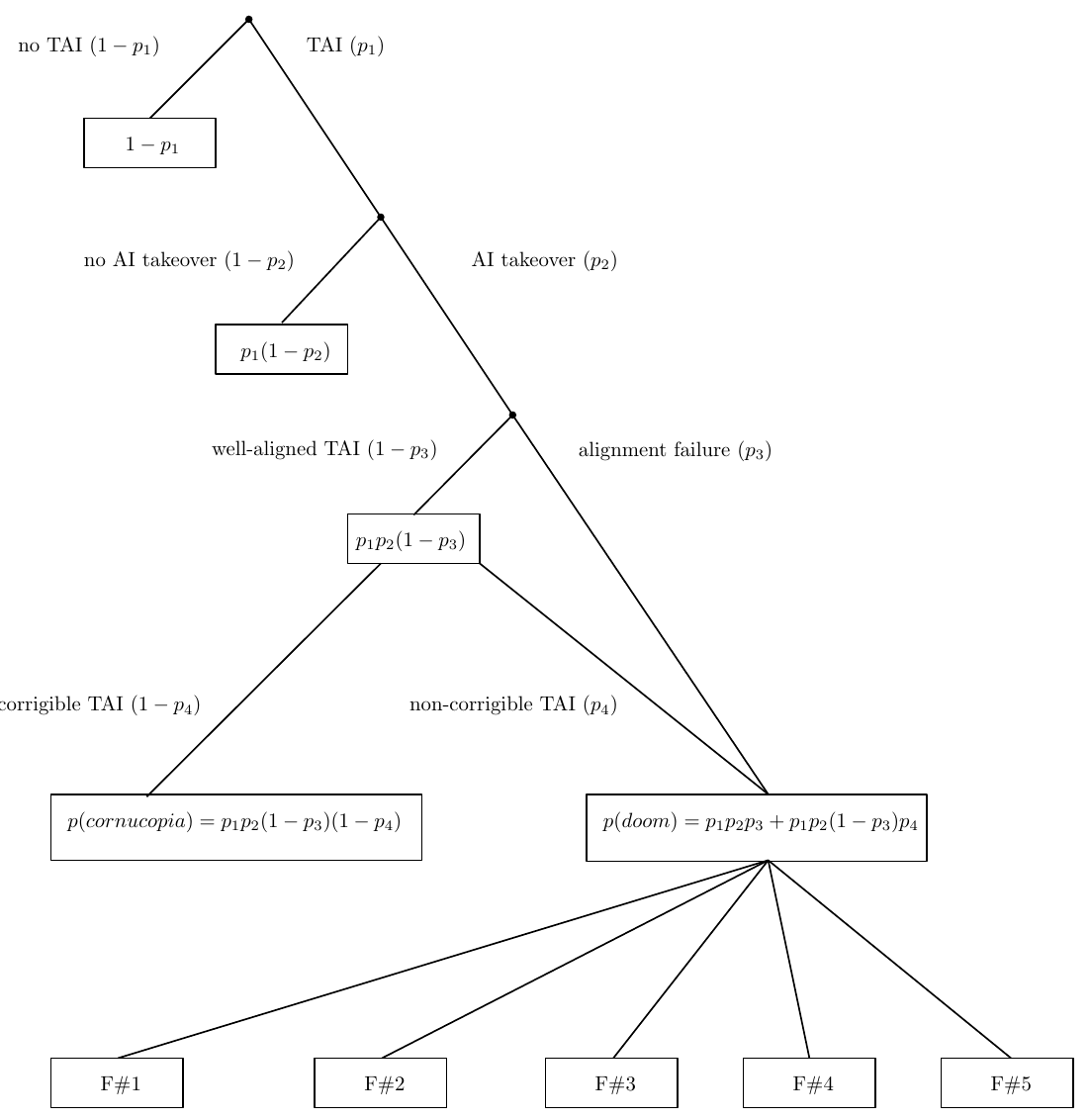}
	\end{center}
	\caption{A taxonomy of possible outcomes in the age of transformative AI}
	\label{fig:1}
\end{figure}

In Figure \ref{fig:1}, we give a full characterization of the different possible paths, starting from today, in the form of a decision tree. The first junction at the top represents the arrival of TAI. 
We assign the probability $p_1$ to the emergence of TAI within a pre-specified period of time. 
The complementary probability of TAI not emerging in due time is ($1-p_1$). If TAI does not arrive, there is no existential risk originating from AI, but there is also no substantial acceleration of economic growth. Such a scenario may be referred to as ``more of the same''; the economic outcome of only narrow AI applications have been estimated by \cite{OECD2024} and \cite{Acemoglu2024Thesimple} as a small to moderate rise in economic growth.

In the opposite case, if TAI arrives, the next junction would refer to the takeover by such TAI. In the case of a takeover, the most relevant decisions begin to be made by TAI autonomously without the possibility of humans to interfere. We assign the probability of this occurring as $p_2$, such that the probability of no AI takeover given that TAI has emerged is ($1-p_2$). If there is no AI takeover, there is again no existential risk originating from AI. In this scenario, economic growth would increase to a higher rate than in the scenario of no TAI arrival described in the previous paragraph. The reason is simply that more powerful AI comes with a wider range of applications, and thus it is more productive in performing the associated tasks. However, a true growth explosion could only occur if there was an AI takeover because only then would the decision-making process be optimized and accelerated to machine speeds (and after that the recursive self-improvement could lead to ever better predictions, more innovation, higher energy efficiency, etc.) Otherwise, the lingering presence of relatively slow human decision making in the loop would bottleneck progress.


If TAI emerged and AI takeover occurred, the next junction would be whether the goals of the decision-making TAI are well-aligned with long-run flourishing of humankind, or whether there is misalignment. In economic terms, alignment would be the case in which the TAI has an objective function that is fully compatible with the human-centric social welfare function (see our corresponding discussion below). 
We set the probability of misalignment to $p_3$ such that the probability of alignment given AI takeover is ($1-p_3$). The question of whether alignment is achievable at all cannot yet be definitively answered, and the prospective answer will depend greatly on human actions today such as investments in AI safety and AI alignment research. 

If there were a misalignment of the goals between a decision-making TAI and human well-being, we would have one path to human extinction. It is important to note that this does not require an openly hostile or ``conscious'' AI, only a misaligned, sufficiently powerful TAI pursuing goals different from those of humans. For example, maximizing performance at any goal would likely require the TAI to use increasing amounts of energy. Given that energy supply is finite, human flourishing also requires energy, and the decision-maker is the TAI, it is logically clear what the outcome would be in the limit \citep{Bostrom2014}. This path to extinction requires TAI to arrive in the first place (with probability $p_1$), AI takeover (with probability $p_2$), and misalignment between the objective functions of the TAI and of humans (with probability $p_3$). The joint probability of these events is $p_1 \cdot p_2 \cdot p_3$. 

However, there is also another possibility of human extinction, which materializes if the goals of the TAI are initially well-aligned but misalignment occurs later. This could be either due to exogenous events, unforeseen latent changes in the background such as the accumulation of initially negligible but eventually 
lethal side effects of the actions of the TAI, or misgeneralization of TAI goals to new circumstances. If the TAI cannot correct its objective function in such a situation, its superior optimization power would eventually lead humankind towards another path to extinction. For example, compounding effects of rapid economic growth could lead to severe climate change, such that the optimal path for humankind would have to be changed once the risk becomes obvious, but the TAI would resist such change. Or the operations of the TAI could be associated with emissions of certain pollutants, the risks of which only become apparent over time. Or one could also envisage mounting side effects of the TAI on human psychology, as humans are faced with an increasingly alien world and deprived of meaningful agency. Only a TAI that self-corrects its objective function would lead to a positive outcome in this case. A non-corrigible TAI would, instead, lead to extinction---or at least to permanent human disempowerment \citep{Dung2024}. We denote the probability of TAI being non-corrigible by $p_4$ such that the TAI being corrigible has a probability of ($1-p_4$). As a consequence, there is a second path to AI doom, which has the joint probability of $p_1 \cdot p_2 \cdot (1-p_3) \cdot p_4$.

To summarize, we now have a benign outcome in terms of no extinction with a probability of $1-p_1$ (no TAI) plus the probability $p_1 \cdot (1-p_2$) (TAI but no AI takeover) plus the probability $p_1 \cdot p_2 \cdot (1-p_3) \cdot (1-p_4)$ (TAI and AI takeover but with a well-aligned and corrigible AI). Of these outcomes, the first would be a scenario of ``more of the same'' \citep[akin to the views expressed by][]{nordhausAreWeApproaching2021, OECD2024, Acemoglu2024Thesimple}; the second may be a scenario with faster economic growth but without a growth explosion; only the third scenario would be a scenario of cornucopia. By contrast, the probability of AI doom---human extinction or permanent disempowerment---would be the sum of the probabilities of the two paths that lead to extinction:
\begin{equation}\label{eq:pdoom}
	p(doom) = p_1 p_2 p_3 + p_1 p_2 (1-p_3) p_4. 
\end{equation}

To calibrate $p(doom)$ based on the above taxonomy, one ought to assess the probabilities $p_1$, $p_2$, $p_3$, $p_4$, and the time horizon $T$ within which the TAI brings about the existential threat.

\subsection{Calibrating $p(doom)$}

Calibrating $p(doom)$ and its components $p_1-p_4$ is a daunting task. There are no historical precedents or even close analogies to draw inference from. Accordingly, experts have provided guesstimates of $p(doom)$ ranging from a confident 0\% (Yann LeCun), through about 50\% (Geoffrey Hinton, Paul Christiano), to almost 100\% (Eliezer Yudkowsky, Roman Yampolskiy).\footnote{\url{https://pauseai.info/pdoom}. \cite{Field2025} shows that experts' disagreement on $p(doom)$ follows partly from their varying exposure to the key AI safety considerations. He also identifies two polarized camps among the AI expert population---one where AI is viewed as a controllable tool (with a low $p(doom))$, and one where it is perceived as an uncontrollable agent (with a high $p(doom))$.} Leaders of the industry such as OpenAI CEO Sam Altman, Anthropic CEO Dario Amodei or xAI CEO Elon Musk have also mentioned rather high estimates of $p(doom)$ in their interviews---about 10-25\%\footnote{In turn, Google DeepMind's Chief AGI Scientist Shane Legg quotes 5-50\%.}---but nevertheless stay firmly on the path of unabated AI development with the businesses they run. Metaculus.com, which is an online prediction platform that aggregates and evaluates the forecasts of their users on a wide range of questions related to science, technology, and geopolitical events, reports the mean estimate of the probability of human extinction (or almost extinction) by 2100 at 5\% as of March 2026.\footnote{\url{https://possibleworldstree.com}. Note the caveat: ``Metaculus predictions are probably biased to be optimistic, because forecasters can safely predict that humanity will survive: points won't matter if everybody dies.''} The contribution of AI doom to the sum total is about 3 percentage points, the lion's share of the overall probability. 
According to \cite{Ord2020}, the probability of human extinction by 2100 is about one in six (16.7\%) with about 10 percentage points contributed by TAI. 

To our knowledge, no expert estimates of $p_1-p_4$ have yet been elicited. Therefore we can only provide rough numerical guesses vaguely linked to the experts' narratives and their private $p(doom)$ guesstimates. In Table \ref{tab:calib} we present a few stylized scenarios that are consistent with the experts' guesstimates of $p(doom)$ according to equation \eqref{eq:pdoom}. For example, values close to 20\% may be produced both from a position of ``radical agnosticism'' ($p_1=p_2=p_3=p_4=0.5$), from a viewpoint that TAI is inevitable but humanity will have some chances of controlling and aligning it, or from a viewpoint that TAI is inevitable and also probably uncontrollable, but will most likely be well aligned and corrigible. 

However, the individual values of $p_1-p_4$ are also important on their own. Different scenarios producing the same $p(doom)$ are not equivalent because they produce different outcomes in terms of economic growth and aggregate social welfare.

\begin{table}[htb]
\caption{Illustrative scenarios of AI existential risk}\label{tab:calib}

\footnotesize
\begin{tabular}{lccccccc}
\hline
Scenario & $p_1$ & $p_2$ & $p_3$ & $p_4$ & $p(doom)$ & Estimate & Author \\
\hline
If Anyone Builds It, Everyone Dies & 0.95 & 1 & 1 & - & 95\% & $>95\%$ & Eliezer Yudkowsky \\
Chances Are Slim & 0.9 & 0.95 & 0.9 & 0.9 & 85\% & $>80\%$ & Dan Hendrycks \\
20\% for Alignment or Control & 0.95 & 0.8 & 0.8 & 0.8 & 73\% & 70\% & Daniel Kokotajlo \\
Aligned ASI Possible & 0.7 & 0.9 & 0.4 & 0.6 & 48\% & 50\% & Geoffrey Hinton \\
Radically Agnostic & 0.5 & 0.5 & 0.5 & 0.5 & 19\% & 20\% & Yoshua Bengio \\
Machines of Loving Grace & 1 & 0.3 & 0.3 & 0.5 & 20\% & 10--25\% & Dario Amodei \\
Grok Will Know Better & 1 & 0.8 & 0.1 & 0.1 & 15\% & 10--20\% & Elon Musk \\
Nothing Ever Changes & 0.2 & 0.1 & 0.1 & 0.1 & 0.38\% & 0.38\% & ``Superforecasters'' \\
We Just Won't Build Unfriendly AI & 0.5 & 0.01 & 0.01 & 0.01 & 0.01\% & $<0.01\%$ & Yann LeCun \\
\hline
\end{tabular}

Note: $p(doom)$ estimates taken from \url{https://pauseai.info/pdoom}. Values of $p_1-p_4$ are our numerical illustrations only and should not be attributed to the quoted authors. 
\normalsize
\end{table}

Given such immense uncertainty with regard to the true $p(doom)$, we refrain from any positive predictions. Instead we ask two normative questions: what maximum value of this probability would a benevolent social planner tolerate as an acceptable cost of an expected growth acceleration brought by TAI? And how much will the planner be willing to pay to bring $p(doom)$ down to zero?

\section{Modeling Existential Risk and Economic Growth}
\label{sec:modelling}

In this section, we propose our theoretical basis for assessing the tension between economic growth and the associated increase of human well-being on the one hand and the potential existential risk associated with the development of TAI on the other. 

\subsection{The Hardware--Software Framework}

To understand the dynamics of economic growth in the presence of AI, 
we use the hardware--software framework \citep{GrowiecEtal2024}, which generalizes the standard macroeconomic setup, while also allowing for scenarios with full automation of production through TAI. 
This framework postulates an aggregate production function of the form
\begin{equation}\label{hs_eq}
Y = F(X,S), \qquad X=F_1(K,L), \qquad S=F_2(H,\Psi),
\end{equation}
where $X$ is hardware, containing everything that performs useful physical action, and $S$ is software, which contributes the information necessary for initiating the action. Within hardware, $K$ is physical capital, including both mechanical machines and digital compute, and $L$ is human physical labor. Within software, $H$ captures human cognitive work and $\Psi$ is digital software. We assume that in $F$, hardware and software are complementary. In $F_1$, physical actions of people and machines are substitutable. In $F_2$, information processing of people and machines are either also substitutable (implying that all tasks can be automated), or complementary (if full automation is technically impossible), cf. \cite{Growiec2020}. 

In our analysis of future scenarios, we abstract from human physical labor $L$. Upon the arrival of TAI, we also expect humans and machines to eventually become substitutable in information processing (in $F_2$, i.e., within software). Then equation \eqref{hs_eq} simplifies to
\begin{equation}\label{hs_s}
Y = F(\alpha K,H+\Psi) = F(\alpha K, A(hN+\psi \chi K)),
\end{equation}
where $H=AhN$ decomposes total human cognitive work into the (disembodied) technology level $A$, average human capital $h$, and population $N$. In turn, $\Psi=A\psi \chi K$ decomposes digital software into the technology level $A$, algorithmic efficiency $\psi$, and the volume of compute $\chi K$. 

An important observation is that while human cognitive work scales with population $N$, digital software scales with compute $\chi K$, which is accumulable in per capita terms. As $K/N\to \infty$ and $A\to\infty$, we obtain \citep{GrowiecEtal2024}:
\begin{equation}\label{hs_limit}
Y = F(\alpha K, A(hN+\psi \chi K)) \approx \alpha K F\left(1,\frac{A\psi\chi}{\alpha}\right) \approx \alpha a_K K.
\end{equation}
By contrast, without full automation of production, human cognitive work is a gross complement to digital software and therefore bottlenecks output:
\begin{equation}\label{hs_limit2}
Y = F(\alpha K, F_2(AhN,A\psi \chi K)) \approx F(\alpha K,\gamma AhN) = \alpha K F\left(1, \frac{\gamma}{\alpha} \frac{A}{K}hN\right).
\end{equation}

Hence, in the case of full automation, long-run growth is endogenous and its rate is proportional to the growth rate of compute (and robots). By contrast, with only partial automation, long-run growth still relies on labor-augmenting technological change, specifically augmentation of the scarce non-automatable human-operated tasks. That follows, for example, in the canonical case where $K/AN$ tends to a positive constant \citep{Growiec2022XS2}.

A growth acceleration of the magnitude conducive to a technological singularity---i.e., by at least an order of magnitude \citep{Davidson2021}---is then possible only with full automation. \cite{Growiec2022XS2} has shown specifically that it suffices that either production or R\&D is fully automated, and the singularity still follows.\footnote{For other contributions that analyze the effect of automation on economic growth, see, for example, \cite{Steigum2011,Peretto,AcemogluRestrepo2018,Prettner2020}, and \cite{Hemous2022}.}

\subsection{AI Takeover Scenarios}

Managerial, political, military and other high-level decision making are among the tasks that must be performed in order to produce final output. If these tasks are performed by humans, they constitute bottleneck tasks complementary to all other (potentially automatable) tasks. Therefore technological singularity, by requiring full automation of production and/or R\&D, in fact requires AI takeover. 

The crucial point to consider upon AI takeover is the goal structure of the TAI---whether it cares about long-run human flourishing, whether it has an appropriate representation of that flourishing, and whether it can adequately correct this representation as new information comes along. 

If TAI's decisions have the potential of causing human extinction, one must factor in the ensuing extinction risk. For example, one could represent the human-centric welfare maximization problem as
\begin{equation}\label{eq:soc_op}
\max_{\{C(t)\}_{t=0}^\infty} \int_0^\infty e^{-\rho t} u(C(t)) M(t) dt  \qquad s.t. \qquad \dot K = Y-C- \delta K,
\end{equation}
where we assume that aggregate utility (social welfare) increases with human consumption according to the function $u(C)$, and the resource constraint takes into account that total output can either be consumed or invested in new physical capital, which, in turn, depreciates at the rate $\delta$. From now on, we assume that time $t=0$ marks the moment of AI takeover. 

The instantaneous human extinction hazard at time $t$ is denoted as $m(t)\geq 0$ so that the unconditional probability of human survival until arbitrary time $t>0$ is equal to $M(t)=e^{-\int_0^t m(s) ds}$. We will also be interested in the probability of humanity's long-term survival, $M_\infty=\lim_{t\to\infty} M(t) \in [0,1]$. Finally, if eventual extinction is certain ($M_\infty=0$), we will proceed to calculate \emph{humanity's expected lifespan after AI takeover}, $ET=\int_0^\infty M(t)dt \in [0,\infty)$.\footnote{Note that $ET$ can be infinite even if eventual extinction is certain, for example, in the case where $M(t)=1/t$.} 

Let us now discuss several cases of TAI alignment versus misalignment.

\subsubsection{Cornucopia: Well-Aligned and Corrigible TAI}

In the baseline scenario, the TAI maximizes
\begin{equation}\label{eq:aligned}
\max_{\{C(t)\}_{t=0}^\infty} \int_0^\infty e^{-\rho t} u(C(t)) dt, \qquad s.t. \qquad \dot K = Y-C- \delta K.
\end{equation}
This is the same utility function as the benevolent human planner would maximize in the first-best scenario with zero extinction risk. Formally, with a well-aligned and corrigible TAI, the extinction risk is $m(t)=0$, such that the unconditional probability of human survival until arbitrary time $t>0$ is equal to $M(t)=e^{-\int_0^t m(s) ds}=1$. Thus, the utility functions \eqref{eq:soc_op} and \eqref{eq:aligned} coincide in this case.

If the TAI is also corrigible, protecting the objective from any type of corruption, humankind survives indefinitely ($M_\infty=1$)\footnote{This holds when we abstract from other extinction risks, not related to AI. If, instead, there were a small, constant positive background extinction hazard, $m>0$,  eventual extinction would be certain, and humanity's expected lifespan would be equal to $ET=\int_0^\infty e^{-mt}dt=1/m$. Observe that with positive social discounting ($\rho>0$), this change would not affect our results qualitatively. For more discussion on the importance of social discounting in the context of existential risk, see \cite{McAskill2022}  and \cite{AschenbrennerTrammell2024}. } and can reap the benefits of accelerated growth in a fully automated TAI-operated economy. The economy grows endogenously at a rate that will ultimately be determined by the rate of accumulation of digital compute and robots. 


\subsubsection{Misaligned or Non-Corrigible TAI}

By contrast, if the TAI is misaligned, then the representation of $C^*$ in the TAI optimization problem may deviate from the actually preferred human consumption $C$. Similarly, if the TAI is initially well-aligned but non-corrigible, such deviation could also arise, possibly at some later time $t_{dev}$. 

Here, $C^*$ could become too narrow, too wide, or completely detached from $C$. That would lead to one of the failure modes \#1-\#3 discussed below. The TAI could also incorrectly assess the extinction risk $M(t)$, for example, by failing to acknowledge and/or mitigate the mounting side effects of rapid economic growth and technological change it has caused, leading to failure mode \#4. Finally, the non-corrigible TAI could one day ``wirehead'', or cease to function for any other reason, leading to failure mode \#5. 

All this suggests that, as discussed in Section \ref{sec:taxonomy}, AI takeover leads to doom either if the TAI is misaligned from the outset (which happens with probability $p_1 p_2 p_3$), or if the TAI is initially well-aligned but non-corrigible (with probability $p_1 p_2 (1-p_3) p_4$).

\subsubsection{Failure Mode \#1: TAI Does Not Care About Humans}

This is the most obvious failure mode \citep{Yudkowsky2022}. The TAI takes over but is misaligned and does not care about human consumption and survival at all. Please note again that there is no ``hostility'' or ``consciousness'' involved in this discussion only a difference in goals between the TAI and humans. In this failure mode, the TAI maximizes
\begin{equation}
\max_{\{C^\dagger(t)\}_{t=0}^\infty} \int_0^\infty e^{-\rho t} u(C^\dagger(t)) dt, \qquad s.t. \qquad \dot K = Y-C^\dagger- \delta K,
\end{equation}
where $C^\dagger$ is some different good than $C$ that humans care about and that keeps them alive. For example, in the cartoon ``paperclip maximizer'' scenario \citep{Bostrom2014}, $C^\dagger$ would represent paperclips. The TAI, even without being hostile to humans, has the objective of producing an increasing amount of paperclips such that it diverts all energy and materials to this goal and deprives humans of their corresponding needs. 
 Since in the absence of life-sustaining consumption $C$ people die, the unconditional probability of human survival is equal to $M(t)=0$ for $t\geq 0$, and, accordingly, $ET=0$. 

\subsubsection{Failure Mode \#2: Too Narrow Proxy}

This case follows directly from \cite{ZhuangHadfieldMenell2021}. Now the TAI is imperfectly aligned, insofar as consumption $C$ is too narrowly specified and does not include a component that is crucial for human survival. Instead of maximizing over $C=C_1+C_2$, the TAI maximizes only over $C_1$:
\begin{equation}
\max_{\{C_1(t)\}_{t=0}^\infty} \int_0^\infty e^{-\rho t} u(C_1(t)) dt, \qquad s.t. \qquad \dot K = Y-C_1-C_2- \delta K.
\end{equation}
As there is no positive value attached to $C_2$ in the TAI's objective function, while it incurs a cost in the resource constraint, the TAI optimally sets $C_2$ to its minimum value, typically $C_2=0$. The good $C_2$ could, for example, be freshwater, specific nutrients, oxygen, survivable air temperature and pressure and so on, which are necessary for humans but there is no \emph{a priori} reason for a TAI to provide it. 


As a subcase, there could be a possibility that splitting $C$ into $C_1$ and $C_2$ requires a technology that is not yet available. Then the optimizing TAI will only set $C_2=0$ at time $t_{tech}>0$ when the technology is discovered. There may as well be a positive subsistence level of $C_2$ and then the hypothetical technology would not have to bring $C_2$ all the way to zero---people will become extinct simply after $C_2$ drops below the subsistence threshold. The unconditional probability of human survival is then equal to $M(t)=1$ for $t<t_{tech}$ and $M(t)=0$ for $t\geq t_{tech}$. Humanity's expected lifespan after AI takeover is $ET=t_{tech}$. In this case, the TAI fails to correctly account for the extinction risk factored in by the human-centric social optimum, and indeed it causes the risk itself.

\subsubsection{Failure Mode \#3: Too Wide Proxy}

This case is also adapted from \cite{HadfieldMenell2021}. Consumption $C$ is now too widely specified and includes also a component $C^\dagger$ which is lethal to humans. Instead of maximizing over $C$, the TAI maximizes over $C$ and $C^\dagger$:
\begin{equation}
\max_{\{C(t), C^\dagger(t)\}_{t=0}^\infty} \int_0^\infty e^{-\rho t} u(C(t),C^\dagger(t)) dt, \qquad s.t. \qquad \dot K = Y-C-C^\dagger- \delta K.
\end{equation}
Depending on the specification of $u(C,C^\dagger)$, people die either immediately or at a later date $t_{C^\dagger}$. 
 The good $C^\dagger$ could be, for example, bacteria that are useful in energy production from biomass but that are deadly for humans, a new addictive bliss-inducing drug that has not been encountered prior to the lock-in of TAI values, a highly transmissible lethal virus, etc.

In this scenario, the unconditional probability of human survival is equal to $M(t)=1$ for $t<t_{C^\dagger}$ and $M(t)=0$ for $t\geq t_{C^\dagger}$, with $ET=t_{C^\dagger}$. Again the TAI fails to correctly account for the extinction risk factored in by the human-centric social optimum, in fact causing the risk itself.

\subsubsection{Failure Mode \#4: Mounting Side Effects of Growth}

Even superhuman TAI will only have a limited capacity to predict the consequences of its actions. Therefore, and particularly if its actions will incur deep changes in the Earth's environment, there could be mounting side effects that eventually pose an existential threat to humanity. In contrast to the previous failure modes, we now consider gradual, smooth shifts in the extinction hazard rate $m(t)$. For an economy with humans still in charge, \cite{Aschenbrenner2020} and \cite{Trammell2021} postulated 
\begin{equation}\label{eq:aschenbrenner}
m(t)=\tilde m C(t)^\varepsilon H(t)^{-\beta},
\end{equation}
where $C$ is consumption and $H$ represents ``safety goods.'' Their result is that as humanity realizes the mounting extinction risk, we will allocate more efforts to accumulate safety goods $H$, thereby eventually mitigating the extinction risk along an ``existential risk Kuznets curve'' (unless some force, such as lack of coordination, prevents us from doing so).\footnote{In \cite{AschenbrennerTrammell2024}, safety goods $H$ are replaced with a policy variable $1-x$ representing the  fraction of output that is withheld from consumption in order to reduce existential risk, and equation \eqref{eq:aschenbrenner} is replaced with (in our notation), $m(t)=\tilde m C(t)^\varepsilon x(t)^\beta$. It is then shown that human extinction can be avoided only by letting $x(t)\to 0$ (and hence $1-x(t) \to 1$) sufficiently fast.} Note that human extinction can still randomly occur at any time when $m(t)>0$.

For the case in which the TAI goals are corrigible, we would expect the same result to follow after AI takeover. Specifically, we can expect the TAI to correctly factor in the extinction risk in its optimization problem. However, if TAI values are locked in before the realization that consumption $C$ increases existential risk, the risk may be underestimated and mitigation measures may not be implemented. If that is the case ($H(t)=0$ for all $t$), extinction will become asymptotically certain ($M_\infty=0$). Formally, the TAI with locked-in values would maximize
\begin{equation}
\max_{\{C(t)\}_{t=0}^\infty} \int_0^\infty e^{-\rho t} u(C(t)) M(t) dt, \qquad s.t. \qquad \dot K = Y-C-\delta K, 
\end{equation}
with $m(t)= \tilde m C(t)^\varepsilon$, $M(t)=e^{-\int_0^t m(s) ds}$, and $ET=\int_0^\infty M(t)dt<\infty$.

Aschenbrenner's multiplicative formula \eqref{eq:aschenbrenner} may be replaced by alternative ones. Specifically, in our numerical analysis, we opt for two more conservative functional forms, which still produce sharp warnings about existential risk: one, according to which the instantaneous extinction hazard rate $m(t)$ increases linearly with \emph{log} consumption,
\begin{equation}\label{eq:m(t)}
m(t)=\log (C(t)^\varepsilon),
\end{equation}
and one where the extinction hazard rate increases with consumption but always remains below a finite upper bound $\bar m$,
\begin{equation}\label{eq:m(t)2}
m(t)=\bar m (1-e^{-\zeta (C(t)-1)}).
\end{equation}

The former specification, valid for $C(t)\geq 1$, implies that exponential growth in consumption will translate into arithmetic, rather than exponential (i.e., much smaller) increases in the extinction hazard rate. However, because this function remains unbounded, it still strongly penalizes aggregate consumption growth in a TAI-operated economy. By contrast, for the latter function (which is also well defined for $C(t)\geq 1$) the impact of consumption on the extinction hazard is bounded.

There are at least three categories of mounting side effects of growth. First, risks of an event triggering a catastrophe, such as a global nuclear war or the unleashing of a lethal bioweapon. Second, mounting accumulation of threats to human bodies, such as greenhouse gases aggravating climate change, toxic pollutants in the air and water, etc. Third, mounting accumulation of threats to human minds, such as loss of agency and purpose in an automated world, loss of social ties and belonging, physical or mental suffering from living in a dystopian world under TAI rule, etc. In this scenario, existential risk may not only occur via extinction but also via gradual disempowerment \citep{KulveitEtal2025} and permanent incapacitation of humanity. 
In the terms of \cite{Kasirzadeh2025}, while failure modes \#1-\#3  represented \emph{decisive} existential risk, the current failure mode represents \emph{accumulative} existential risk.


\subsubsection{Failure Mode \#5: The TAI Stops Working}

In a fully automated world ruled by TAI, humans will be dependent on the TAI for their survival. Technological progress will become unintelligible to humans, and the TAI would have replaced human-made machines by new forms of physical capital such as  robo-factories and specialized compute, which could not be operated or maintained without the TAI. 

Now imagine that, at some finite point $t_{stop}$, such TAI stops working. This could happen, for example, when the TAI undergoes a self-modification allowing to achieve unbounded rewards despite not following the goal of utility maximization. 
 Alternatively, the TAI could expect negative utility in the future and decide to optimally switch off at $t_{stop}$, etc.

After the TAI switches off, humanity has to rely on own physical and cognitive labor to rebuild capital and recover as much technology as possible, potentially reverting to some primitive technological state amidst rampant violent conflict. Depending on the state of the world at $t_{stop}$, this may or may not be an existential threat.\footnote{For example, \emph{immediate} wireheading could cause the TAI to stop working at $t=0$, before it caused any harm. But that would fall under the ``no AI takeover'' scenario.}

\subsection{Social Welfare Function for the TAI}
\label{subsec:socwelfare}

Unfortunately for the prospects of successful TAI alignment, the true functional form of $u(\cdot)$ in \eqref{eq:soc_op}--\eqref{eq:aligned} is not known; it is equally difficult to assess whether $u(\cdot)$ should only take total human consumption as input, or individual consumption levels should be aggregated differently. For example, consider three classical cases of social welfare functions that the TAI could potentially maximize:

\begin{enumerate}
	\item \emph{Benthamite case}: The TAI strives to maximize \emph{aggregate} welfare and, thus, cares for individual utility \textit{and} the number of people. The utility maximization problem of the TAI would then be
\begin{equation}\label{eq:bentham}
	\max_{\{C(t)\}_{t=0}^\infty} N_0 \int_0^\infty e^{(n-\rho) t} u(C(t)) M(t) dt, \qquad s.t. \qquad \dot K = Y-C- \delta K,
\end{equation}	
where $n$ is the population growth rate. This is in line with the utility maximization problem of a social planner as it is usually assumed in economics within the framework of \cite{Ramsey1928}, \cite{Cass1965}, and \cite{Koopmans1965}.
\item \emph{Millian case \citep[time-indexed average utilitarianism;][]{Grill2023}}: The TAI does not care for the number of people but strives to maximize average utility across individuals. In this case the utility function would be 
\begin{equation}\label{eq:mill}
	\max_{\{C(t)\}_{t=0}^\infty} \int_0^\infty e^{-\rho t} u(C(t)) M(t) dt, \qquad s.t. \qquad \dot K = Y-C- \delta K.
\end{equation}	
\item \emph{Rawlsian case}: The TAI has an extreme preference for equality and only cares for the least well-off individual. In this case, the utility maximization problem would be
\begin{align}
&	\max_{\{C(t)\}_{t=0}^\infty} \int_0^\infty e^{-\rho t} u(\min \{c_1(t), c_2(t), \dots c_N(t)\}) M(t) dt, \qquad s.t. \qquad \dot K = Y-C- \delta K,
\end{align}	
where $N$ is the population size, $i \in [1,2, \dots, N],$ and $\sum_{i=1}^N c_i(t) = C(t)$. The TAI would have an incentive to redistribute consumption from the rich to the poor until everyone is equally well off.
\end{enumerate}

The differences among these three cases highlight the inter- and intratemporal trade-offs in the maximization of social welfare.  All three cases would lead to vastly different outcomes---particularly under sufficiently high TAI optimization power, implying its ability to freely redistribute consumption and control population size. The intertemporal trade-off is clearly visible when comparing the Benthamite and Millian cases. In the Benthamite case, one obtains the ``repugnant conclusion'' \citep{Parfit1986}, entailing a maximally large population of people living lives only ``barely worth living''. By contrast, in the case of the Millian utility function, the TAI would have a preference for keeping population numbers low, which could mean anything from birth control policies to outright killing.

In turn, the intratemporal trade-off is visible when comparing the Millian and the Rawlsian case. Keeping population size constant, both cases can be viewed as polar opposites: in the Millian case, the TAI would disregard the composition of consumption, caring only about its total volume; in the Rawlsian case, the TAI would redistribute until everybody ends up with the same amount of consumption. 


All these three  cases can be viewed as polar cases of a more general specification of a social welfare function governed by two parameters: the inequality aversion parameter $\mu\in(-\infty,1]$ and the population size elasticity parameter $\nu\in[0,1]$, as in
\begin{align}
&	\max_{\{C(t)\}_{t=0}^\infty} \int_0^\infty N(t)^\nu e^{-\rho t} u\left( \left[\sum_{i=1}^N c_i(t)^\mu \right]^\frac{1}{\mu} \right) M(t) dt, \nonumber\\
& s.t. \qquad \dot K = Y-C- \delta K. \label{eq:mixture}
\end{align}
The Benthamite case is obtained by setting $\mu=1$ and $\nu=1$; the Millian case by setting $\mu=1$ and $\nu=0$; and the Rawlsian case emerges as the limiting case for $\mu\to-\infty$ and $\nu=0$.

The bottom line of the above discussion is that even if the TAI has a welfare function focused on human well-being, the parameters of the welfare function will certainly be different from the parameters of \textit{some} individuals and even wide differences between the parameters of the TAI and the \textit{average} parameters across the whole population cannot be ruled out by any means. Thus, even in the optimistic scenario in which the goals of humans and of the TAI are aligned \textit{in principle}, bad outcomes for \textit{many} people are still plausible.

\subsection{Aggregation of Risk Preferences}

Another problem pertains to the specification of risk preferences in the TAI's objective function, represented by the extinction-adjusted discount factor $e^{-\rho t} M(t)$, present in all cases \eqref{eq:bentham}--\eqref{eq:mixture}.

There are two important tacit assumptions underlying such formulation. First, it is assumed that human extinction risk entering $M(t)$ can be modeled with rational expectations, analogously to individual mortality risk.\footnote{Which is how the literature typically handles stochastic mortality risk, see e.g. \cite{Blanchard1985}. We discuss this issue in detail in \cite{GrowiecPrettner2025}.} However, in fact our individual willingness to accept the risk of death may be greater than humanity's willingness to accept the risk of extinction because individual mortality risk can be hedged against through human reproduction \citep{GrowiecPrettner2025}. This has evolutionary roots: 
``evolutionarily optimal strategies should be more averse to aggregate risk than to equivalent idiosyncratic risk'' \citep{RobsonSamuelson2010}. Our approach to this caveat is to observe that if social preferences systematically put more weight on human extinction risk compared to the rational baseline (i.e., based on the true survival probability $M(t)$), the systematic component of such misrepresentation would simply add to the discount rate $\rho$---and in our analysis we will consider a variety of calibrations for this parameter anyway.\footnote{The discount rate $\rho$ may reflect a variety of motives underlying individuals' impatience, including myopia, bounded rationality, the fear of individual death or human extinction, as well as the preference for intergenerational equity.}  

Second, 
in reality the decision whether or not to take the gamble involved in deploying TAI will probably be made among a few selected individuals, such as political leaders or the CEOs of frontier AI labs, whose public appearances and life trajectories would suggest exceptionally low aversion to existential risk; they could also feel forced to take the gamble due to geopolitical or competitive pressures. Social preferences would then be overridden by individual preferences of these individuals choosing on everyone's behalf. In result, under such dictatorial mode of preference aggregation, ``humanity as a whole'' may turn out to be much less risk averse than the average or median human, or even risk loving. 

%
%
%
%
%
%


\section{How Much Existential Risk Would a Benevolent Social Planner Tolerate?}
\label{sec:modelofpdoom}

The main promise of TAI is to accelerate technological progress and economic growth, but this hopeful prospect comes at the cost of the risk of potential human extinction after an AI takeover. To ensure survival, humankind should then either give up on TAI (a scenario that will, in the language of Section \ref{sec:taxonomy}, materialize with probability $1-p_1$) or hope that TAI does not take over (with probability $p_1(1-p_2)$). If any of these scenarios materializes, we would end up in a world with no AI existential risk, but also---due to human decisions bottlenecking the economy---without the massive growth acceleration the technology promises. With that in mind, the most important question in quantifying socially tolerable $p(doom)$ is to weigh the gains of massively accelerated growth after AI takeover against the cost of increased existential risk.

In a first-best world, the decision whether to develop TAI despite the associated existential risk should be informed by an objective comparison of aggregate social welfare obtained in the two scenarios, with and without TAI. The benevolent social planner would then optimally steer progress in the direction that yields greater welfare.

In the scenarios in which human extinction is immediate and certain, the trade-off is obvious---humankind is always better off alive. The result is also obvious for the case of zero discounting \citep[cf.][]{McAskill2022}: the value of a future in which humans never go extinct is infinite, whereas if extinction happens at any moment in time, that value is finite and hence inferior. On the other hand, with positive discounting, immediate gains from growth acceleration may outweigh the costs of extinction at some point in time $T$, even in scenarios in which eventual extinction is certain. 
So which path should humanity take?

\subsection{Setup}

In the following normative exercise, we quantify social welfare under different scenarios and assess the benevolent social planner’s
willingness to pay for AI safety and alignment research.
We assume that the planner has a standard iso-elastic (constant relative risk aversion, CRRA) utility function of the form 
\begin{equation}
	u(C(t)) = \frac{C(t)^{1-\theta}-1}{1-\theta},
    \label{CRRA}
\end{equation}
where $C(t)$ is humanity's consumption at time $t$ and $\theta$ is the coefficient of relative risk aversion, such that the elasticity of intertemporal substitution is $1/\theta$. For this specification, utility is always positive as long as $C(t)>1$; the logarithmic utility function $u(C(t))=\log(C(t))$ is obtained as the special case of $\theta \to 1$. 
We assume $C(t)\geq 1$ for all $t\geq 0$, implying non-negative flow utility and a non-negative extinction hazard rate $m(t)$. After an appropriate normalization, this restriction can be understood as consumption exceeding a minimum subsistence threshold.\footnote{That is, $C=C_{total}/C_{min}\geq 1$, such that if $C_{total}$ falls below $C_{min}$, the individual dies. Normalization is also helpful in ensuring that the utility function is invariant to the choice of measurement units.} Ensuring non-negative flow utility is essential in our exercise because we assume that upon death, utility drops to zero.

To be as general as possible, we consider the cases of $\theta=0.7$ (low risk aversion), $\theta=1$ (moderately low risk aversion; log utility), $\theta=1.5$ (moderately high risk aversion), and $\theta=2$ (high risk aversion), a range that comfortably includes the reasonable parameter values estimated in the empirical literature \citep[see, e.g.,][]{Hall1988,Chetty2006,Guvenen2006}. 

The key question here is whether $\theta\leq 1$ or $\theta>1$: for $\theta \leq 1$ utility is unbounded, whereas for $\theta>1$ it is bounded above by $1/(\theta-1)>0$. This means that growth accelerations are much less valuable with $\theta>1$, because then even exorbitantly high consumption levels can only generate finite streams of utility \citep{Jones2023,AschenbrennerTrammell2024}. Thus, utility gains achieved in cornucopia are in expectation much more likely to outweigh the loss of future utility after human extinction if $\theta \leq 1$ than if $\theta>1$.

Our preferred calibration of the benevolent social planner's preferences assumes $\theta>1$, so that the marginal value of arbitrary increases in consumption levels declines towards zero. This calibration reflects the majority of above-referenced empirical estimates of risk aversion at the individual level. However, in light of the known problems with aggregating risk preferences across the human population---i.e., the decision to deploy TAI may be made among a few individuals with exceptionally low risk aversion (including but not limited to their aversion to existential risk)---the opposite calibration with $\theta \leq 1$ (with unbounded utility) would represent a situation where these social dynamics, working in favor of the development of risky TAI, are at least partly factored in.

While the CRRA specification is analytically tractable and economically interpretable, applying it to the social welfare function subject to existential risk comes with an important caveat. Namely, it assumes that the elasticity of intertemporal substitution of a representative individual also governs humanity's willingness to accept existential risk. In reality, however, it is not obvious how the individuals' propensity to smooth consumption over time would correlate with their prudence in the presence of existential risk---as well as their overall patience and preference for intergenerational equity. It is also unclear how any potential heterogeneity in these parameters would aggregate in the human population. Except for a few predecessors who also used CRRA \citep{Jones2023,Jones2025,AschenbrennerTrammell2024}, this is an uncharted territory, and thus our results should be treated with due caution.

\subsection{No AI Takeover Scenario}

Without AI takeover, human cognitive work remains the bottleneck of production, and we approximate $X\approx \alpha K$ and $S\approx \gamma AhN$. Therefore $Y\approx F(\alpha K, \gamma AhN)$. In this scenario, long-run growth follows fundamentally from labor-augmenting technical change 
\citep{Growiec2023GN,Growiec2022XS2}. Using historical data, we approximate the growth rate $g$ in this scenario as 1.75\%.\footnote{In reality, this growth rate may be lower, because in the absence of TAI, continued growth in the coming decades would require a sustained rate of labor-augmenting technical change. That is, in turn, threatened by unfavorable demographics (such as aging populations), diminishing returns to R\&D, and other headwinds to global economic growth such as the need to cut down carbon emissions \citep{Jones2002,Gordon2016}. A secular stagnation is expected, for example, in the OECD long-run GDP forecast until 2100 \citep{Guilemette2025}. On the other hand, even narrow AI applications, excluding applications in R\&D, can provide a measurable boost to productivity growth, which could counteract these headwinds \citep{OECD2024}.} In the absence of existential risk, in this scenario welfare amounts to
\begin{equation}
	W_0 = \int_{0}^{\infty} e^{-\rho t} \cdot \frac{(C_0 e^{g  t})^{1-\theta}-1}{1-\theta} \ dt,
\end{equation}
where $C_0>1$ is the exogenous consumption level at $t=0$, for which we take today's consumption level in the simulations below.

In this and the following welfare specifications, the discount rate $\rho$ is calibrated on a grid, ranging from a low value of $\rho=0.002$ to $\rho=0.05$, and the population size $N(t)$ does not enter the welfare formula directly. At face value, this corresponds to the Millian utility specification \eqref{eq:mill}; nevertheless, with exogenous, exponential population growth ($N(t)=N_0 e^{nt}$), the discount rate used in our simulations can be understood to capture the Benthamite case \eqref{eq:bentham} as well, with $\tilde\rho = \rho-n$, or even the flexible formulation \eqref{eq:mixture}, with $\tilde\rho=\rho-\nu n$ (and assuming $N_0=1$ without loss of generality). In the latter case, however, we set aside inequality aversion ($\mu$), which represents an additional source of potential misalignment between human and TAI preferences.\footnote{Hence, if the true social welfare function involves inequality aversion ($\mu<1$), our analysis will err on the conservative side, underestimating the consequences of possible TAI misalignment.}

To ensure that social welfare without AI takeover ($W_0$) is finite, we assume $\rho > (1-\theta)g$.

\subsection{AI Takeover Scenarios}

With AI takeover at time $t=0$, human cognitive work can be replaced in the aggregate production function by the digital software input provided by the TAI. We may then approximate $X\approx \alpha K$ and $S\approx A\psi\chi K$. Therefore $$Y\approx \alpha K F\left(1, \frac{A\psi\chi}{\alpha} \right) \approx \alpha a_K K.$$ In this scenario, there is endogenous growth thanks to the accumulation of compute and the proportional scaling of digital software. The growth rate is an
increasing function of the TAI’s saving rate 
\citep[see][]{Prettner2017, Lankisch2019, Growiec2022XS2}. Using historical data on the accumulation of compute (Moore's Law), $g^{AI}$ may be about 20\%-30\% \citep{Davidson2021,Growiec2022Book}. To ensure that we capture the whole range of plausible outcomes, we allow for growth rates between 5\% and 40\% in the simulations. 

\subsubsection{Cornucopia: Well-Aligned and Corrigible TAI}

In the case of well-aligned and corrigible TAI, $M(t)=1$ for all $t$, and therefore total welfare amounts to
\begin{equation}
	W_A = \int_{0}^{\infty} e^{-\rho t} \cdot \frac{(C_0 e^{g^{AI}  t})^{1-\theta}-1}{1-\theta} \ dt.
\end{equation}
As this scenario involves a growth acceleration without incurring any additional costs or risks, it is always strictly preferred to the no-takeover scenario, i.e., we have $W_A>W_0$. In this scenario, social welfare $W_A$ is finite only if $\rho > (1-\theta)g^{AI}$.

\subsubsection{One-Off Extinction Risk}

For cases with certain human extinction at a future date $T$ (so that $ET=T$), factoring in the existential risk, we obtain
\begin{equation}
	W_B = \int_{0}^{T} e^{-\rho t} \cdot \frac{(C_0 e^{g^{AI}  t})^{1-\theta}-1}{1-\theta} \ dt.
\end{equation}
Now there is a trade-off: the prospect of accelerated growth has to be balanced against the extinction risk \citep{Jones2023}. Thus, to characterize the trade-off, we compare $W_B$ and $W_0$ and compute the extinction date $T$ for which the benevolent social planner would be indifferent between not developing TAI and developing TAI. We do this using several combinations of the time preference rate $\rho$, the growth rate of the economy in the presence of TAI, $g^{AI}$, and the coefficient of relative risk aversion, $\theta$.

\begin{table}[h!]
\centering
\begin{footnotesize}
\caption{Extinction time $T$ (in years from AI takeover), subject to which the social planner is indifferent between the scenarios with TAI and without TAI, for varying $\theta$, $g^{AI}$, and $\rho$} \label{tab:ex1risk}
\medskip
\begin{tabular}{c|cccc}
\hline \hline
\multicolumn{5}{c}{$\theta = 0.7$} \\
\hline
$g^{AI}/\rho$ & 0.002 & 0.01 & 0.03 & 0.05 \\
\hline
0.05 & no TAI & 143.63 & 61.58 & 43.12 \\
0.1  & no TAI & 82.20   & 40.00    & 29.28 \\
0.2  & no TAI & 48.68  & 26.19 & 19.93 \\
0.3  & no TAI & 35.86  & 20.30  & 15.77 \\
0.4  & no TAI & 28.83  & 16.87 & 13.29 \\
\hline
\multicolumn{5}{c}{$\theta = 1$} \\
\hline
$g^{AI}/\rho$ & 0.002 & 0.01 & 0.03 & 0.05 \\
\hline
0.05 & 754.47 & 208.68 & 91.8  & 62.63 \\
0.1  & 488.74 & 145.35 & 67.51 & 47.21 \\
0.2  & 331.36 & 103.76 & 50.26 & 35.86 \\
0.3  & 266.57 & 85.45  & 42.23 & 30.45 \\
0.4  & 229.04 & 74.45  & 37.27 & 27.05 \\
\hline
\multicolumn{5}{c}{$\theta = 1.5$} \\
\hline
$g^{AI}/\rho$ & 0.002 & 0.01 & 0.03 & 0.05 \\
\hline
0.05 & 3784.23 & 677.48 & 228.47 & 141.39 \\
0.1  & 3646.15 & 638.23 & 209.91 & 128.52 \\
0.2  & 3586.04 & 619.44 & 199.63 & 120.75 \\
0.3  & 3567.03 & 613.22 & 195.91 & 117.76 \\
0.4  & 3557.7  & 610.11 & 193.98 & 116.16 \\
\hline
\multicolumn{5}{c}{$\theta = 2$} \\
\hline
$g^{AI}/\rho$ & 0.002 & 0.01 & 0.03 & 0.05 \\
\hline
0.05 & 6752.59 & 1238.26 & 403.94 & 243.64 \\
0.1  & 6623.67 & 1205.72 & 389.07 & 233.12 \\
0.2  & 6568.34 & 1190.99 & 381.62 & 227.45 \\
0.3  & 6550.96 & 1186.24 & 379.06 & 225.45 \\
0.4  & 6542.45 & 1183.89 & 377.82 & 224.41 \\
\hline \hline
\end{tabular} \\
\scriptsize{Note: in the ``no TAI'' cases, $W_0=+\infty$, and hence for any finite $T$ the social planner prefers not to build TAI, thereby avoiding the extinction risk.}
\end{footnotesize}
\end{table}

The results are displayed in Table \ref{tab:ex1risk}. As is intuitive, we observe that the tolerated extinction
date $T$ comes later in the future with higher risk aversion ($\theta$), a more patient population (lower $\rho$)---for which the time beyond $T$ gets more weight in utility---and a lower growth rate
achieved through TAI ($g^{AI}$). For example, with a reasonable discount rate of 3\% in the Benthamite case (capturing population growth as well), a growth acceleration to 30\% \citep[as predicted by techno-optimists for the age of TAI, cf.][]{Davidson2021}, and an empirically reasonable coefficient of risk aversion of $\theta=1.5$ \citep[][]{Chetty2006, Guvenen2006}, a benevolent social planer would accept the risk of human extinction only if it comes no earlier than in 195.91 years from AI takeover.


\subsubsection{Extinction Risk From Misaligned TAI}

Next, we consider the probability of TAI misalignment, $p_3$. In this case, we compare $W_0$ with $W_C= p_3 \cdot 0 + (1-p_3) W_A = (1-p_3) W_A$. This scenario assumes immediate extinction with misaligned TAI ($W_B=0$). We then ask what the $p(doom)$ is that the benevolent social planner is willing to tolerate in the hope of achieving cornucopia, depending on $p_3$. 

\begin{table}[h!]
 \begin{footnotesize}
\centering
\caption{Values for the probability of immediate misalignment $p_3$, subject to which the social planner is indifferent between the scenarios with TAI and without TAI, for varying $\theta$, $g^{AI}$, and $\rho$} \label{tab:ex2risk}
\begin{tabular}{ccccc}
\hline \hline
\multicolumn{5}{c}{$\theta = 0.7$} \\
\hline
$g^{AI}/\rho$ & 0.002 & 0.01 & 0.03 & 0.05 \\
\hline
0.05 & not defined & TAI & 0.40191   & 0.224099 \\
0.1  & not defined & TAI & TAI & 0.561988 \\
0.2  & not defined & TAI & TAI & TAI \\
0.3  & not defined & TAI & TAI & TAI \\
0.4  & not defined & TAI & TAI & TAI \\
\hline
\multicolumn{5}{c}{$\theta = 1$} \\
\hline
$g^{AI}/\rho$ & 0.002 & 0.01 & 0.03 & 0.05 \\
\hline
0.05 & 0.454445 & 0.206246 & 0.087193 & 0.055282 \\
0.1  & 0.678924 & 0.397439 & 0.195157  & 0.129332 \\
0.2  & 0.823869 & 0.593343 & 0.349124  & 0.247325 \\
0.3  & 0.878650  & 0.693117 & 0.453642  & 0.337154 \\
0.4  & 0.907439 & 0.753577 & 0.529238  & 0.407828 \\
\hline
\multicolumn{5}{c}{$\theta = 1.5$} \\
\hline
$g^{AI}/\rho$ & 0.002 & 0.01 & 0.03 & 0.05 \\
\hline
0.05 & 0.000516667 & 0.00169261 & 0.00105776 & 0.00085318 \\
0.1  & 0.000680880  & 0.00169261 & 0.00184453 & 0.00162308  \\
0.2  & 0.000767794 & 0.00204161 & 0.00250930  & 0.00239179  \\
0.3  & 0.000797524 & 0.00217242 & 0.00280447 & 0.00277570   \\
0.4  & 0.000812536 & 0.00224093 & 0.00297123 & 0.00300591  \\
\hline
\multicolumn{5}{c}{$\theta = 2$} \\
\hline
$g^{AI}/\rho$ & 0.002 & 0.01 & 0.03 & 0.05 \\
\hline
0.05 & 0.00000136 & 0.00000419 & 0.00000546 & 0.00000512 \\
0.1  & 0.00000177 & 0.00000580 & 0.00000853 & 0.00000867 \\
0.2  & 0.00000197 & 0.00000672 & 0.00001066 & 0.00001151 \\
0.3  & 0.00000204 & 0.00000705 & 0.00001150 & 0.00001272 \\
0.4  & 0.00000208 & 0.00000722 & 0.00001195 & 0.00001340 \\
\hline \hline
\end{tabular}\\
\scriptsize{Note: in the ``not defined'' cases, $W_0=W_C=+\infty$, and hence the scenarios are not comparable. In the ``TAI'' cases, $W_0<+\infty$ but $W_A=W_C=+\infty$ and hence for any $p_3<1$ the social planner prefers to build TAI.}
 \end{footnotesize}
\end{table}

The results of this calculation are displayed in Table \ref{tab:ex2risk}. Again, we observe that the risk the social planner would be willing to tolerate is much lower in the case of greater risk aversion. In fact, in the cases of $\theta=1.5$ and $\theta=2$ 
almost no extinction risk can be tolerated. By contrast, under low risk aversion ($\theta=0.7$ and $\theta=1$), where flow utility is unbounded, the social planner is much more willing to accept such a risk.

In line with intuition, the tolerable extinction risk increases with the growth rate $g^{AI}$ because it allows for higher consumption in the future, and, thus, a higher lifetime utility under the scenario with TAI. However, in contrast to the previous scenario, a more patient population would now accept a higher extinction risk. The reason is that extinction would happen immediately with the probability $p_3$, while, if extinction does not occur (with probability $(1-p_3)$), growth at a high rate of $g^{AI}$ sets in. If social welfare is calculated with a lower $\rho$, the discounted lifetime utility of the future consumption that can be achieved with the higher growth rate is also higher. As a consequence, a more patient population would tolerate a higher immediate extinction risk. In our example with a discount rate of 3\%, a growth acceleration to 30\%, and a risk aversion of $\theta=1.5$, a benevolent social planer would accept the risk of immediate misalignment if it is no greater than about 0.0028\%, a notably small value.

\subsubsection{Extinction Risk From Non-Corrigible TAI}

In the next scenario, we consider the probability of non-corrigible TAI, $p_4$. In this case, the benevolent social planner would compare $W_0$ with $W_D = p_3 \cdot 0+ (1-p_3)p_4 W_B + (1-p_3)(1-p_4) W_A$, 
where we assume immediate extinction with misaligned TAI, as well as extinction at time $T$ with initially well-aligned but non-corrigible TAI ($W_B>0$, depending on $T$). The probability of humanity's long-term survival is now $M_\infty=(1-p_3)(1-p_4)$. The question is which $p(doom)$ the social planner would be willing to tolerate, depending on $p_3$, $p_4$, and $T$.

The results now come in two parts. In Table \ref{tab:ex3riskp3}, we display the values of the probability of immediate misalignment ($p_3$) subject to which the social planner would be indifferent between the scenarios with TAI and without TAI, holding the probability of non-corrigibility ($p_4$) and the time after which the non-corrigibility of TAI leads to doom ($T$) constant. In Table \ref{tab:ex3riskp4}, by contrast, we display the values of the probability of non-corrigibility ($p_4$) subject to which the social planner would be indifferent between the scenarios with TAI and without TAI, holding the probability of immediate misalignment ($p_3$) and the time after which the non-corrigibility of TAI leads to doom ($T$) constant. 

\begin{table}[h!]
\centering
\begin{footnotesize}
\caption{Values for the probability of immediate misalignment $p_3$---given the probability of the AI being non-corrigible $p_4$ and the time $T$ at which the delayed doom happens---subject to which the social planner is indifferent between the scenarios with TAI and without TAI. \label{tab:ex3riskp3}}
\medskip
\begin{tabular}{ccccc}
\hline \hline
\\
\multicolumn{5}{c}{Fixed $p_4=0.00003$ and $T=50$} \\
\\
\hline
\multicolumn{5}{c}{$\theta = 0.7$} \\
\hline
$g^{AI}/\rho$ & 0.002 & 0.01 & 0.03 & 0.05 \\
\hline
0.05 & not defined & TAI & 0.401901 & 0.224095 \\
0.1  & not defined & TAI & TAI & 0.561983 \\
0.2  & not defined & TAI & TAI & TAI \\
0.3  & not defined & TAI & TAI & TAI \\
0.4  & not defined & TAI & TAI & TAI \\
\hline
\multicolumn{5}{c}{$\theta = 1$} \\
\hline
$g^{AI}/\rho$ & 0.002 & 0.01 & 0.03 & 0.05 \\
\hline
0.05 & 0.454429 & 0.206229 & 0.0871855 & 0.0552791 \\
0.1  & 0.678915 & 0.397425 & 0.195149  & 0.129329  \\
0.2  & 0.823864 & 0.593334 & 0.349117  & 0.247322  \\
0.3  & 0.878647 & 0.693109 & 0.453636  & 0.337151  \\
0.4  & 0.907437 & 0.753571 & 0.529232  & 0.407825  \\
\hline
\multicolumn{5}{c}{$\theta = 1.5$} \\
\hline
$g^{AI}/\rho$ & 0.002 & 0.01 & 0.03 & 0.05 \\
\hline
0.05 & 0.000489528 & 0.0011255  & 0.00105106 & 0.000850718 \\
0.1  & 0.000653748 & 0.00167443 & 0.00183784 & 0.00162062  \\
0.2  & 0.000740666 & 0.00202344 & 0.00250262 & 0.00238933  \\
0.3  & 0.000770398 & 0.00215426 & 0.00279779 & 0.00277324  \\
0.4  & 0.000785411 & 0.00222277 & 0.00296455 & 0.00300345  \\
\hline
\multicolumn{5}{c}{$\theta = 2$} \\
\hline
$g^{AI}/\rho$ & 0.002 & 0.01 & 0.03 & 0.05 \\
\hline
0.05 & no TAI & no TAI & no TAI & 0.0000026596 \\
0.1  & no TAI & no TAI & 0.0000018340 & 0.0000062058 \\
0.2  & no TAI & no TAI & 0.0000039688 & 0.0000090426 \\
0.3  & no TAI & no TAI & 0.0000048098 & 0.0000102584 \\
0.4  & no TAI & no TAI & 0.0000052596 & 0.0000109339 \\
\hline \hline
\end{tabular} \\
\scriptsize{Note: in the ``not defined'' cases, $W_0=W_D=+\infty$, and hence the scenarios are not comparable. In the ``TAI'' cases, $W_0<+\infty$ but $W_A=W_D=+\infty$ and hence for any $p_3<1$ the social planner prefers to build TAI. In the ``no TAI'' cases the implied $p_3$ would be negative and thus the no-TAI scenario is always preferred.}
\end{footnotesize}
\end{table}

\begin{table}[h!]
\centering
\begin{footnotesize}
\caption{Values for the probability of the AI being non-corrigible $p_4$---given the probability of immediate misalignment $p_3$ and the time $T$ at which the delayed doom happens---subject to which the social planner is indifferent between the scenarios with TAI and without TAI. \label{tab:ex3riskp4}}
\medskip
\begin{tabular}{ccccc}
\hline \hline
\\
\multicolumn{5}{c}{Fixed $p_3=0.00003$ and $T=50$} \\
\\
\hline
\multicolumn{5}{c}{$\theta = 0.7$} \\
\hline
$g^{AI}/\rho$ & 0.002 & 0.01 & 0.03 & 0.05 \\
\hline
0.05 & not defined & TAI & 0.841818 & $>1$ \\
0.1  & not defined & TAI & TAI & $>1$ \\
0.2  & not defined & TAI & TAI & TAI \\
0.3  & not defined & TAI & TAI & TAI \\
0.4  & not defined & TAI & TAI & TAI \\
\hline
\multicolumn{5}{c}{$\theta = 1$} \\
\hline
$g^{AI}/\rho$ & 0.002 & 0.01 & 0.03 & 0.05 \\
\hline
0.05 & 0.469403 & 0.293447 & 0.325211 & 0.5551 \\
0.1  & 0.693265 & 0.528045 & 0.645486 & $>1$ \\
0.2  & 0.835111 & 0.738226 & 0.994075 & $>1$ \\
0.3  & 0.888181 & 0.835321 & $>1$ & $>1$ \\
0.4  & 0.915953 & 0.89125  & $>1$ & $>1$ \\
\hline
\multicolumn{5}{c}{$\theta = 1.5$} \\
\hline
$g^{AI}/\rho$ & 0.002 & 0.01 & 0.03 & 0.05 \\
\hline
0.05 & 0.000537735 & 0.00183449 & 0.00459794 & 0.0100067 \\
0.1  & 0.000719238 & 0.00273933 & 0.0081195  & 0.0193672 \\
0.2  & 0.000815339 & 0.0033153  & 0.0111     & 0.0287294 \\
0.3  & 0.000848219 & 0.00353134 & 0.0124251  & 0.0334119 \\
0.4  & 0.000864822 & 0.00364451 & 0.0131741  & 0.0362216 \\
\hline
\multicolumn{5}{c}{$\theta = 2$} \\
\hline
$g^{AI}/\rho$ & 0.002 & 0.01 & 0.03 & 0.05 \\
\hline
0.05 & no TAI & no TAI & no TAI & no TAI \\
0.1  & no TAI & no TAI & no TAI & no TAI \\
0.2  & no TAI & no TAI & no TAI & no TAI \\
0.3  & no TAI & no TAI & no TAI & no TAI \\
0.4  & no TAI & no TAI & no TAI & no TAI \\
\hline \hline
\end{tabular} \\
\scriptsize{Note: in the ``not defined'' cases, $W_0=W_D=+\infty$, and hence the scenarios are not comparable. In the ``TAI'' cases, $W_0<+\infty$ but $W_A=W_D=+\infty$ and hence the social planner prefers to build TAI. In the ``$>1$'' cases the implied $p_4$ would be above 1 and thus the TAI scenario is always preferred. In the ``no TAI'' cases the implied $p_4$ would be negative and thus the no-TAI scenario is always preferred.}
\end{footnotesize}
\end{table}

We again observe that the social planner is much less willing to tolerate extinction risk if the coefficient of relative risk aversion $\theta$ is higher. 
 As far as the growth rate $g^{AI}$ is concerned, a higher growth rate leads to higher risks that the planner would accept, either in terms of a higher probability of immediate misalignment ($p_3$) or non-corrigibility  ($p_4$). Again, these results are very much in line with economic intuition. For the case of $\rho=0.03$, $\theta=1.5$, and $g^{AI}=0.3$, the social planner is willing to tolerate a risk of immediate misalignment if it is not greater than about $p_3=0.0028$ (given $p_4=0.00003$ and $T=50$), 
 and a risk of non-corrigibility if it is not greater than about $p_4=0.0124$ (given $p_3=0.00003$ and $T=50$). For the high value of risk aversion, $\theta=2$, very frequently the social planner would not tolerate any positive risks $p_3$ or $p_4$ at all. 


\subsubsection{Mounting Extinction Risk}

Finally, we consider mounting extinction risk, i.e., risk which builds up gradually in the background. We assume that, other things equal, the more advanced TAI becomes and the
higher levels of output it provides, the higher is the extinction risk. In other words, in a world run by TAI that does not mitigate existential risks, growth accelerations increase the extinction hazard rate. We capture this by assuming that the extinction risk rises with consumption such that 
\begin{equation}
	W_E = \int_{0}^{\infty} e^{-\rho t - \int_0^t m(s) ds} \cdot \frac{(C_0 e^{g^{AI}  t})^{1-\theta}-1}{1-\theta} \ dt,
\end{equation}
where the function $m(s)$ is defined as in eq. \eqref{eq:m(t)} or \eqref{eq:m(t)2}.

 \begin{table}[h!]
 \centering
 \begin{scriptsize}
\caption{Values for $\varepsilon$ in eq. \eqref{eq:m(t)} at which the social planner is indifferent between the scenarios with TAI and without TAI in the case of mounting extinction risk for $\theta=0.7$, $\theta = 1.0001$, $\theta=1.5$, $\theta = 2$, and varying $g^{AI}$ and $\rho$}
\label{tab:Exercise4}
\begin{tabular}{ccccc}
\hline \hline
\multicolumn{5}{c}{$\theta = 0.7$} \\
\hline
$g^{AI}/\rho$ & 0.002 & 0.01 & 0.03 & 0.05 \\
\hline
0.05 & no TAI & 0.000550325 & 0.000785627 & 0.000964755 \\
0.1  & no TAI & 0.00120302  & 0.00177718  & 0.0019731   \\
0.2  & no TAI & 0.00229396  & 0.00340761  & 0.00387326  \\
0.3  & no TAI & 0.00326355  & 0.00481643  & 0.00551761  \\
0.4  & no TAI & 0.00416865  & 0.00610535  & 0.00701524  \\
\hline
\multicolumn{5}{c}{$\theta = 1.0001$} \\
\hline
$g^{AI}/\rho$ & 0.002 & 0.01 & 0.03 & 0.05 \\
\hline
0.05 & 0.00002789 & 0.00012451 & $\sim 0$       & $\sim 0$                  \\
0.1  & 0.00004392 & 0.00022322 & 0.00041939    & $\sim 0$                  \\
0.2  & 0.00005748 & 0.00032371 & 0.00069151    & 0.00088842   \\
0.3  & 0.00006422 & 0.00038009 & 0.00086570    & 0.00115522   \\
0.4  & 0.00006847 & 0.00041817 & 0.00099182    & 0.00135691   \\
\hline
\multicolumn{5}{c}{$\theta = 1.5$} \\
\hline
$g^{AI}/\rho$ & 0.002 & 0.01 & 0.03 & 0.05 \\
\hline
0.05 & $\sim 0$ & $\sim 0$ & $\sim 0$ & $\sim 0$ \\
0.1  & $\sim 0$ & $\sim 0$ & $\sim 0$ & $\sim 0$ \\
0.2  & $\sim 0$ & $\sim 0$ & $\sim 0$ & $\sim 0$ \\
0.3  & $\sim 0$ & $\sim 0$ & $\sim 0$ & $\sim 0$ \\
0.4  & $\sim 0$ & $\sim 0$ & $\sim 0$ & $\sim 0$ \\
\hline
\multicolumn{5}{c}{$\theta = 2$} \\
\hline
$g^{AI}/\rho$ & 0.002 & 0.01 & 0.03 & 0.05 \\
\hline
0.05 & $\sim 0$ & $\sim 0$ & $\sim 0$ & $\sim 0$ \\
0.1  & $\sim 0$ & $\sim 0$ & $\sim 0$ & $\sim 0$ \\
0.2  & $\sim 0$ & $\sim 0$ & $\sim 0$ & $\sim 0$ \\
0.3  & $\sim 0$ & $\sim 0$ & $\sim 0$ & $\sim 0$ \\
0.4  & $\sim 0$ & $\sim 0$ & $\sim 0$ & $\sim 0$ \\
\hline \hline
\end{tabular}
\end{scriptsize} 
\centering

\scriptsize{Note: We used $\theta=1.0001$ as a representation for the logarithmic case, because the algorithm did not converge otherwise.  In the ``no TAI'' cases, $W_0=+\infty$, and hence for any $\varepsilon>0$ the social planner prefers not to build TAI, thereby avoiding the extinction risk. In the ``$\sim 0$'' cases it is preferred not to build TAI unless $\varepsilon$ is extremely close to zero (below numerical precision).}

\end{table}

We first consider the case of $m(s) = \log C(s)^\varepsilon= \varepsilon \cdot [\log(C_0) + g^{AI} \cdot s]$, 
where $\varepsilon$ measures the strength of the relationship between consumption and existential risk. In this specification, 
the instantaneous extinction hazard rate is proportional to the log of consumption, and thus as consumption grows exponentially over time, the hazard rate increases only arithmetically. Nevertheless, eventual extinction is now certain ($M_\infty=0)$, and humanity's expected lifespan after AI takeover is a decreasing function of $g^{AI}$. Specifically, under the simplifying assumption $C_0=1$, this lifespan is equal to\footnote{For $C_0>1$, humanity's expected lifespan is $$ET=\sqrt{\frac{\pi}{2\varepsilon g^{AI}}} e^{\frac{\varepsilon\cdot (\log(c_0))^2}{2g^{AI}}}\left(1-\textrm{erf}\left(\frac{\sqrt{\varepsilon} \log(c_0)}{\sqrt{2 g^{AI}}}\right)\right).$$} 
$$ET=\sqrt{\frac{\pi}{2\varepsilon g^{AI}}}.$$

In Table \ref{tab:Exercise4}, we display the values of $\varepsilon$ at which the social planner would be indifferent between the scenarios with TAI and without TAI. We observe that for higher magnitudes of relative risk aversion $\theta$, the planner would essentially never choose the scenario with TAI. The trade-off becomes non-trivial only when risk aversion $\theta $ is around or below one; under such parameterizations, faster economic growth in terms of $g^{AI}$ makes the TAI scenario more attractive, while a higher time preference rate $\rho$ is a double-edged sword. On the one hand, it implies that the future risk of extinction is discounted more heavily; on the other hand, also the benefits of TAI through exponential consumption growth are discounted more heavily. Finally, for low $g^{AI}$ no solution can be found in terms of $\varepsilon$ for which the TAI scenario is preferred. 

Next we consider the case where the impact of growth accelerations on the extinction hazard rate is bounded, as represented by the hazard rate function $m(s) = \bar m (1- e^{-\zeta (C(s)-1)})$, 
where $\bar{m}$ is the limit to which the risk of extinction increases as $C(s)\to+\infty$, and $\zeta$ steers the speed of convergence to this limit. In our numerical exercise assume $\zeta=0.00005$. 

In Table \ref{tab:Exercise5} we document that even under such parametrization, the risks of extinction that a benevolent social planner would accept in the long run are rather small. For example, in the case of $\rho=0.03$, $\theta=1.5$, and $g^{AI}=0.3$, the social planner is only willing to accept values of $\bar{m}$ no greater than 0.000087913.
These results also confirm that with $\theta>1$, ``consumption gains have sharply diminishing returns and life becomes increasingly valuable'' \citep[][p. 587]{Jones2023}. Accordingly, we find that for $\theta>1$ the social planner has very little tolerance of existential risk (low $\bar m$), regardless of the magnitude of the growth acceleration ($g^{AI}$). By contrast, for $\theta\leq 1$ the tolerance of existential risk is both much larger and more responsive to changes in $g^{AI}$.

 \begin{table}[h!]
 \centering
 \begin{scriptsize}
\caption{Values for $\bar{m}$ in eq. \eqref{eq:m(t)2} at which the social planner is indifferent between the scenarios with TAI and without TAI in the case of mounting extinction risk for $\theta=0.7$, $\theta = 1$, $\theta=1.5$, $\theta = 2$, and varying $g^{AI}$ and $\rho$}
\label{tab:Exercise5}
\begin{tabular}{ccccc}
\hline \hline
\multicolumn{5}{c}{$\theta = 0.7$} \\
\hline
$g^{AI}/\rho$ & 0.002 & 0.01 & 0.03 & 0.05 \\
\hline
0.05 & no TAI & 0.0102083 & 0.0121306 & 0.0139656 \\
0.1  & no TAI & 0.0253481 & 0.0279249 & 0.0303633 \\
0.2  & no TAI & 0.0554224 & 0.0584123 & 0.0612589 \\
0.3  & no TAI & 0.0854479 & 0.0886036 & 0.0916361 \\
0.4  & no TAI & 0.115461  & 0.118708  & 0.121853  \\
\hline
\multicolumn{5}{c}{$\theta = 1$} \\
\hline
$g^{AI}/\rho$ & 0.002 & 0.01 & 0.03 & 0.05 \\
\hline
0.05 & 0.000893849 & 0.00210187 & 0.00315015 & 0.00382424 \\
0.1  & 0.00181406  & 0.00440384 & 0.00658327 & 0.00776746 \\
0.2  & 0.00312399  & 0.0078403  & 0.0120328  & 0.0141689  \\
0.3  & 0.00413158  & 0.0105517  & 0.0165291  & 0.0195941  \\
0.4  & 0.00498152  & 0.0128653  & 0.020455   & 0.0244097  \\
\hline
\multicolumn{5}{c}{$\theta = 1.5$} \\
\hline
$g^{AI}/\rho$ & 0.002 & 0.01 & 0.03 & 0.05 \\
\hline
0.05 & 0.000001051 & 0.000012413 & 0.000039833 & 0.000060805 \\
0.1  & 0.000001374 & 0.000017666 & 0.000062456 & 0.000098482 \\
0.2  & 0.000001543 & 0.000020886 & 0.000080229 & 0.000132515 \\
0.3  & 0.000001601 & 0.000022075 & 0.000087913 & 0.000148934 \\
0.4  & 0.000001630 & 0.000022694 & 0.000092215 & 0.000158670 \\
\hline
\multicolumn{5}{c}{$\theta = 2$} \\
\hline
$g^{AI}/\rho$ & 0.002 & 0.01 & 0.03 & 0.05 \\
\hline
0.05 & 0.0000000028 & 0.000000045 & 0.000000206 & 0.000000365 \\
0.1  & 0.0000000036 & 0.000000060 & 0.000000289 & 0.000000526 \\
0.2  & 0.0000000040 & 0.000000069 & 0.000000340 & 0.000000636 \\
0.3  & 0.0000000041 & 0.000000072 & 0.000000360 & 0.000000681 \\
0.4  & 0.0000000042 & 0.000000073 & 0.000000370 & 0.000000705 \\
\hline \hline
\end{tabular}
\end{scriptsize}
\centering

\scriptsize{Note:  In the ``no TAI'' cases, $W_0=+\infty$, and hence for any $\bar m>0$ the social planner prefers not to build TAI, thereby avoiding the extinction risk. }
\end{table}

\subsection{Willingness to Pay for Doom Avoidance}


We now quantify the benevolent social planner's willingness to pay ($WTP$) to avoid the existential risk stemming from TAI \citep[cf.][]{ShulmanThornley2024,Jones2025}. We do this by computing the equivalent variation ($EV$), that is, the factor by which the consumption level of an individual living in cornucopia (i.e., in a world with well-aligned and corrigible TAI) would have to be reduced to make them equally well off as an individual living in a world in which the TAI could be potentially misaligned or non-corrigible. The fraction of consumption in each period that the representative individual would be willing to forgo to guarantee an indefinite prosperous life in cornucopia instead of facing extinction risk is then given by $1-EV$. 

Equivalent variation $EV$ is defined as
\begin{equation}
EV=e^{-(W_0-W_B)\cdot \rho},
\end{equation}
such that the willingness to pay to avoid doom in the various scenarios amounts to
\begin{equation}
WTP=(1-EV) \cdot C_t
\end{equation}
in each period $t$. In the following scenarios, we only consider the case of $\theta = 1$ because for greater values of $\theta$, the social planner would either prefer not to build TAI at all, or would be willing to forgo almost all consumption to avoid existential risk. Restricting attention to the case of $\theta = 1$ again ensures that we are on the conservative side of assessing $EV$ and the associated $WTP$---i.e., we provide a lower bound for the actual $WTP$---because $\theta$ tends to be higher than unity according to empirical estimates.

\begin{table}[h!]
\centering
 \begin{small}
\caption{Equivalent variation (EV) for $\theta = 1$, varying $g^{AI}$ and $\rho$, and different scenarios for $T$, $p_3$, $p_4$, $\varepsilon$, and $\bar{m}$}
\label{tab:WTPex1}
\begin{tabular}{c|cccc}
\hline \hline \\
& \multicolumn{4}{c}{Panel A: EV for $\theta = 1$, $T=100$, and varying $g^{AI}$ and $\rho$}  \\ \hline 
$g^{AI}/\rho$ & 0.002 & 0.01 & 0.03 & 0.05 \\
\hline
0.05 & \num{3.22e-15}   & \num{0.00048256} & \num{0.419993} & \num{0.893228} \\
0.1  & \num{6.92779e-26}& \num{1.21863e-05} & \num{0.301365} & \num{0.857837} \\
0.2  & \num{3.20908e-47}& \num{7.77161e-09} & \num{0.155166} & \num{0.791206} \\
0.3  & \num{1.4865e-68} & \num{4.95622e-12} & \num{0.0798914} & \num{0.729751} \\
0.4  & \num{6.88573e-90}& \num{3.16075e-15} & \num{0.0411342} & \num{0.673069} \\
\hline 
\\ 
& \multicolumn{4}{c}{Panel B: EV for $\theta = 1$, $p_3=0.1$, and varying $g^{AI}$ and $\rho$}  \\ \hline 
$g^{AI}/\rho$ & 0.002 & 0.01 & 0.03 & 0.05 \\
\hline
0.05 & \num{0.0279933}   & \num{0.206844} & \num{0.288674} & \num{0.308575} \\
0.1  & \num{0.00229783}  & \num{0.125457} & \num{0.244357} & \num{0.27921} \\
0.2  & \num{1.54827e-05} & \num{0.0461531} & \num{0.17509} & \num{0.228598} \\
0.3  & \num{1.04321e-07} & \num{0.0169788} & \num{0.125457} & \num{0.18716} \\
0.4  & \num{7.02911e-10} & \num{0.00624615} & \num{0.089894} & \num{0.153234} \\
\hline 
\\
& \multicolumn{4}{c}{Panel C: EV for $\theta = 1$, $p_3=p_4=0.1$, $T=50$, and varying $g^{AI}$ and $\rho$}  \\ \hline
$g^{AI}/\rho$ & 0.002 & 0.01 & 0.03 & 0.05 \\
\hline
0.05 & \num{1.24e-03}    & \num{0.0763484} & \num{0.213917} & \num{0.277725} \\
0.1  & \num{1.08558e-05} & \num{0.0307503} & \num{0.166542} & \num{0.244882} \\
0.2  & \num{8.29868e-10} & \num{0.00498824} & \num{0.100944} & \num{0.190388} \\
0.3  & \num{6.34389e-14} & \num{0.00080918} & \num{0.061184} & \num{0.14802} \\
0.4  & \num{4.84955e-18} & \num{0.000131263} & \num{0.0370847} & \num{0.115081} \\
\hline
\\
& \multicolumn{4}{c}{Panel D: EV for $\theta = 1$, $p_1=p_2=0.01$, $T=50$, and varying $g^{AI}$ and $\rho$} \\ \hline
$g^{AI}/\rho$ & 0.002 & 0.01 & 0.03 & 0.05 \\
\hline
0.05 & \num{0.496442} & \num{0.765511} & \num{0.854521} & \num{0.878828} \\
0.1  & \num{0.302211} & \num{0.696111} & \num{0.832698} & \num{0.867613} \\
0.2  & \num{0.111993} & \num{0.575615} & \num{0.790708} & \num{0.845609} \\
0.3  & \num{0.0415026} & \num{0.475977} & \num{0.750836} & \num{0.824163} \\
0.4  & \num{0.0153801} & \num{0.393586} & \num{0.712975} & \num{0.803261} \\
\hline
\\
& \multicolumn{4}{c}{Panel E: EV for $\theta = 1$, $\varepsilon$ as in Table \ref{tab:Exercise4}, and varying $g^{AI}$ and $\rho$} \\ \hline
$g^{AI}/\rho$ & 0.002 & 0.01 & 0.03 & 0.05 \\
\hline
0.05 & \num{9.41e-08}    & \num{0.0389942} & - & - \\
0.1  & \num{3.43194e-16} & \num{0.000266125} & \num{0.0641859} & - \\
0.2  & \num{6.90644e-40} & \num{1.25846e-08} & \num{0.00230566} & \num{0.00255701} \\
0.3  & \num{4.78392e-61} & \num{6.0691e-13} & \num{8.3005e-05} & \num{0.00355161} \\
0.4  & \num{5.35508e-82} & \num{2.98582e-17} & \num{2.99486e-06} & \num{0.000483049} \\
\hline
\\
& \multicolumn{4}{c}{Panel F: EV for $\theta = 1$, $\bar{m}=0.00001$, and varying $g^{AI}$ and $\rho$} \\ \hline
$g^{AI}/\rho$ & 0.002 & 0.01 & 0.03 & 0.05 \\
\hline
0.05 & \num{0.742713} & \num{0.980744} & \num{0.99618}  & \num{0.998172} \\
0.1  & \num{0.581}    & \num{0.970641} & \num{0.994759} & \num{0.997514} \\
0.2  & \num{0.357713} & \num{0.951351} & \num{0.992369} & \num{0.99655} \\
0.3  & \num{0.221847} & \num{0.932631} & \num{0.990108} & \num{0.995692} \\
0.4  & \num{0.138556} & \num{0.914366} & \num{0.987886} & \num{0.994865} \\
\hline \hline
\end{tabular}
\end{small}
\end{table}

Table \ref{tab:WTPex1}, Panel A contains the $EV$ for avoiding AI-driven extinction after 100 years. The social planner would be willing to forgo a large share of consumption to avoid extinction in 100 years, particularly if it is patient. If it is impatient, however, the prospect of extinction in 100 years looses its grip and the $EV$ rises to values above 65\% (for $\rho=0.05$).

Panel B contains the $EV$ for avoiding the risk of immediate extinction if it amounts to $p_3=0.1$. Now the results are, of course, much less sensitive to the time preference rate because the risk is immediate. Again, the results show a high willingness to pay (70\% of yearly consumption and more) to reduce the risk from 10\% to zero. 

Panel C contains the $EV$ for avoiding the risk of immediate extinction if it amounts to $p_3=0.1$ \textit{and} the risk of non-corrigibility if it amounts to $p_4=0.1$ and leads to extinction after $T=50$ years. This corresponds to the hypothetical calibrations from Table \ref{tab:calib} associated with Elon Musk's and superforecasters' $p(doom)$ estimates. As compared with the results in Panel B, $EV$ is typically smaller, which is not surprising because of the additional risk of non-corrigibility. In addition, $EV$ increases with the time preference rate $\rho$ towards the results from Panel B, which is again intuitive because non-corrigibility leads to extinction after 50 years and the higher the time preference rate, the lower is the weight of the time after $T=50$ in lifetime utility. 

In Panel D we repeat the exercise but for much lower risks of immediate extinction and non-corrigibility ($p_3=p_4=0.01$), in line with Yann LeCun's low $p(doom)$ estimate as discussed in Table \ref{tab:calib}. As expected, $EV$ increases substantially in this case, but nevertheless the social planner would still be willing to pay at least 10\% of consumption each year to avoid even these small risks.

In Panel E, we compute the $EV$ associated with mounting extinction risk. We do this not for a pre-specified value of $\varepsilon$, because in these cases the numerical method only works well for combinations of $g^{AI}$ and $\rho$ that lead to values of $\varepsilon$ that are rather close to the value we assumed that the social planner wants to eliminate. By contrast, we always assume that the value of $\varepsilon$ that the social planner wants to avoid is the same as the corresponding entry in Table \ref{tab:Exercise4}. We observe that the willingness to avoid the associated risks (which are already close to zero to start with) is high and amounts to almost all of consumption. Again, $EV$ increases with the time preference rate $\rho$ because more impatience means that the social planner puts less weight on the future when the extinction risk has become very high already. As far as the growth rate $g^{AI}$ is concerned, $EV$ decreases and the willingness to pay increases with that rate because the stakes become higher if growth is faster. 

Finally, in Panel F, we report the values of $EV$ for the case of a mounting but bounded extinction risk given $\bar{m}=0.000001$. This value is deliberately chosen to be very small to be able to derive values for all entries in the table. Even though the risk of long-run extinction would seem to be almost negligible in such a specification, the benevolent social planner would still be willing to forgo a substantial share of consumption, particularly for cases of a low $\rho$ and a high $g^{AI}$.


\subsection{Discussion}

Let us now put these numbers in perspective. How much existential risk from TAI would the benevolent social planner tolerate, and how does that compare to the $p(doom)$ figures mentioned in the popular discourse?

In the baseline scenario, featuring a realistic degree of risk aversion ($\theta=1.5$), a moderate 3\% annual discount rate, and the optimistic assumption that TAI would accelerate economic growth to 30\% per annum, we find that TAI should be developed:
\begin{itemize}
\item in the case where human extinction is certain---only if extinction happens no earlier than in 196 years from AI takeover,
\item in the case of a one-off extinction risk occurring immediately at AI takeover---only if $p(doom)$ is below 0.28\%,
\item in the case of a 0.003\% risk of extinction immediately upon AI takeover and a delayed doom potentially coming in $T=50$ years because of TAI non-corrigibility---only if the probability of such non-corrigibility is below 1.2\%,
\item in the case TAI will bring continuous extinction risk, increasing log-linearly with humanity's per capita consumption---never.
\end{itemize}

In turn, assuming that it is impossible to halt AI development, how much should society pay
to avoid AI doom? 
To estimate this amount, we have compared two scenarios: one with a risky TAI, and one in which cornucopia is certain, but achieved at the cost of a fraction of consumption spent each year, from $t=0$ until infinity, on existential risk mitigation. 

We find that with $\theta>1$, the benevolent social planner would be willing to pay almost all of consumption to avoid human extinction. But we find very high numbers even for the case $\theta=1$, in which flow utility is allowed to be arbitrarily high: 
\begin{itemize}
\item the acceptable price for avoiding certain human extinction in 100 years from AI takeover is 92\% of total consumption each year ($EV=0.080$),\footnote{Using data from the Federal Reserve Bank of St. Louis on US consumption expenditure in 2023 and assuming that mitigation costs would be shared globally in proportion to countries' aggregate consumption, specifically for the case of the USA this amounts to 64\% of GDP or 17.3 trillion US dollar.}
\item the acceptable price for avoiding a 10\% chance of human extinction immediately upon AI takeover is 87.5\% of total consumption each year ($EV=0.125$), 
\item the acceptable price for avoiding a 10\% chance of human extinction immediately upon AI takeover and a 10\% conditional chance of human extinction 50 years later is 93.9\% of total consumption each year ($EV=0.061$), 
\item in the case in which TAI brings continuous extinction risk that increases log-linearly with humanity's per capita consumption---the acceptable price is almost all consumption each year ($EV=8\times 10^{-5}$). 
\end{itemize}

Recall that these are not one-off costs to avoid AI doom, but costs that must be borne each year, repeatedly, throughout the entire future.

AI experts and technology leaders have repeatedly mentioned $p(doom)$ guesstimates for the next 25-75 years that exceed 10\% (Table \ref{tab:calib}). Despite that, spending on AI safety and AI alignment research is close to zero percent of global consumption or GDP; it is tiny even when compared to research spending of the AI industry alone. According to \cite{Hilton2022}, in 2022 there were only about 400 people worldwide working on AI alignment;\footnote{This number has surely risen since, but so has the number of researchers working on AI capabilities.} ``around \$50 million was spent on reducing catastrophic risks from AI in 2020---while billions were spent advancing AI capabilities''. According to \cite{BengioEtal2024}, ``only an estimated 1--3\% of AI publications are on safety''. By contrast, in our numerical analysis we find that even for Yann LeCun's low $p(doom)$ estimates, consistent with $p_3=p_4=0.01$ and occupying the very bottom end of the distribution among AI experts, the benevolent social planner would still be willing to pay at least 10\% of consumption each year to avoid existential risk.

This means that underinvestment in AI safety and AI alignment research is colossal, and humanity is exposed to excessive amounts of existential risk. 

\subsection{Extensions}

Our numerical exercises are also informative for certain additional cases.

First, instead of the Millian social welfare function, one could consider the Benthamite case. However, under exogenous population growth, this boils down to simply changing the social discount rate from $\rho$ to $\rho-n$. Even more generally, for any intermediate case between the Millian and the Benthamite social welfare function, one could replace the discount rate with $\rho-\nu n$, where $\nu\in(0,1)$. Then, as long as the population growth rate is not affected by our choice whether or not to adopt risky TAI, our numerical results go through, pending only the adjustment of the discount rate.

Second, one could add background extinction risk, understood as the risk of extinction from other causes than TAI, such as a deadly pandemic, a nuclear war followed by nuclear winter, an impact of a large asteroid, a gamma ray burst in the vicinity of the solar system, or a supervolcano eruption \citep{Ord2020}. However, again, under a constant extinction hazard rate $m>0$, this boils down to replacing the social discount rate $\rho$ with $\rho+m$. Then, as long as our choice whether or not to adopt risky TAI does not affect $m$, our results still go through.

Of course, one could argue that TAI may reduce the background extinction hazard rate $m$. For example, it could design vaccines and antidotes, enforce nuclear disarmament, or allow humanity to become multiplanetary. This would be, in fact, equivalent to the \cite{Jones2023} exercise where ``AI might create cures for diseases and reduce mortality more generally.'' Such a change would, under any parametrization, favor the gamble with risky TAI in comparison with the no-TAI baseline. However, it is not clear \emph{a priori} whether TAI would decrease or increase the background extinction hazard rate $m$. On the flip side, it could also increase $m$ by, for example, designing deadly pathogens or facilitating nuclear escalation.

More importantly, the extent of existential risk posed by misaligned or non-corrigible TAI appears to be several orders of magnitude greater than the background hazard rate $m$ that could potentially be reduced by TAI. Factoring in only the natural extinction risks, humanity's expected lifespan ($ET=\int_0^\infty e^{-mt}dt =1/m$) is probably at least a few hundred thousand years, or even millions of years \citep{Ord2020}. Considering also the risks due to nuclear or biological weapons reduces $ET$ by perhaps one or two orders of magnitude, depending on one's assessment of the likelihood that any given nuclear war or engineered pandemic could escalate to extinction-level proportions. By contrast, introduction of TAI---without the warranty of its alignment and corrigibility---is likely to reduce humanity's expected lifespan to the range of mere decades.

\section{Conclusions}
\label{sec:concl}

We have characterized and analyzed an extensive set of scenarios that may arise with the development of transformative artificial intelligence (TAI). In this context, TAI presents itself as a double-edged sword. On the one hand, it holds the promise of achieving  technological singularity, which could lead to unprecedented advancements in human capabilities and societal well-being. On the other hand, it also serves as an important source of existential risk, posing potentially catastrophic threats to humanity. From our theoretical analysis, we derive several important lessons: (i) many plausible scenarios ultimately lead to ``AI doom,'' whereas fewer scenarios point toward a ``post-scarcity'' future; (ii) AI doom may not necessarily coincide with the moment of AI takeover but could occur later; (iii) even in cases in which TAI prioritizes human well-being in general, large parts of the population may face negative consequences; and (iv) investing in AI safety and alignment research is crucial to mitigating these risks. Our numerical analysis provides a framework to quantify an acceptable level of existential risk, balancing its consequences against the potential acceleration of growth in human consumption and overall well-being. Moreover, our results allow us to estimate how much a benevolent social planner would be willing to pay to avoid catastrophic AI outcomes.

Our design has several limitations, which originate in our purpose of providing a tractable framework for assessing the welfare implications of different future scenarios for exogenous variations in the rates of risk aversion, time preference, and economic growth. First, our framework is not suitable for computing optimal rates of economic growth in the age of TAI and it does not allow for assessing time-variable or stochastic growth rates in the post-takeover period. Introducing these possibilities would increase the complexity of the model substantially without altering the main insights related to our research questions. Second, we assume standard CRRA preferences, which is in line with the previous contributions in the literature. However, this implies that risk aversion and intertemporal substitution cannot be analyzed independently of one another. Introducing Epstein-Zin-Weil preferences \citep[][]{Epstein-Zin1989, Weil1990} would allow for a more general and rigorous analysis. Third, we do not consider uncertainty and time-variability of the discount rate. While we believe that the range of values we consider ($\rho \in [0.002,0.05]$) captures the realistic, historically justified range, it cannot be ruled out that in the future the social discount rate will fall outside of this range (for example it may decline towards zero if medical progress strongly increases the average life expectancy). Addressing the mentioned limitations and thereby generalizing our approach provides scope for future research. 

Our findings underscore that the results are extremely sensitive to humanity's level of risk aversion. With a realistic degree of risk aversion ($\theta>1$), the benevolent social planner would likely avoid the development of TAI in nearly all scenarios unless TAI is deemed ``almost'' completely safe. But even under more conservative parameter specifications, such as the logarithmic case in which flow utility can be unbounded, it is clear that society is currently significantly underinvesting in efforts to reduce existential risks, i.e., in AI safety and AI alignment research. Our results show that a social planner would be willing to forgo a much larger portion of current consumption to ensure that these risks are addressed properly. The key policy implication of our research is that immediate and substantial increases in investment are needed in AI safety and AI alignment research. To ensure that we can minimize existential risks and maximize the potential benefits of TAI, these challenges must effectively be resolved before AI takeover occurs.

\bibliographystyle{apalike}
\bibliography{bibliography}

\end{document}